\documentclass{article}

\usepackage{arxiv}

\usepackage{amsmath,amssymb,amsfonts}
\usepackage{textcomp}
\usepackage{algorithmic}
\usepackage{siunitx}
\usepackage{algorithmic}
\usepackage{xcolor}

\usepackage{hyperref}

\usepackage{graphicx}
\usepackage{subfig}
\usepackage{multirow}
\usepackage{dblfloatfix}

\begin{document}
%
\title{COVIDGR dataset and COVID-SDNet methodology  for predicting COVID-19 based on Chest X-Ray images}
%
%
%


\author{
  S. Tabik\thanks{Corresponding author} \\
  Andalusian Research Institute\\ in Data Science and\\ Computational Intelligence \\ University of Granada\\ 18071 
  Spain\\ 
  \texttt{siham@ugr.es} \\
  \And
  A. Gómez-Ríos\\
  Andalusian Research Institute\\ in Data Science and\\ Computational Intelligence \\ University of Granada\\ 18071 
  Spain\\ 
  \And
  J.L. Martín-Rodríguez\\Hospital Universitario Clínico \\ San Cecilio de Granada, Spain\\  
  \And
   I. Sevillano-García\\ 
  Andalusian Research Institute\\ in Data Science and\\ Computational Intelligence \\ University of Granada\\ 18071 
  Spain\\ 
 \And
   M. Rey-Area\\
   atlanTTic Research Center for \\ Telecommunication Technologies\\ University of Vigo\\ Galicia, Spain\\
   \And
    D. Charte\\
  Andalusian Research Institute\\ in Data Science and\\ Computational Intelligence \\ University of Granada\\ 18071 
  Spain\\ 
  \And
    E. Guirado\\ Multidisciplinary Institute for \\ Environment Studies “Ramón Margalef”\\ University of Alicante\\ 03690, Spain\\
    \And
    J.L. Suárez\\
  Andalusian Research Institute\\ in Data Science and\\ Computational Intelligence \\ University of Granada\\ 18071 
  Spain\\ \And
    J. Luengo\\
  Andalusian Research Institute\\ in Data Science and\\ Computational Intelligence \\ University of Granada\\ 18071 
  Spain\\ \And
    M.A. Valero-González\\ Hospital Universitario Clínico \\ San Cecilio de Granada, Spain
    \And P. García-Villanova\\ Hospital Universitario Clínico \\San Cecilio de Granada, Spain
    \And E. Olmedo-Sánchez\\ Hospital Universitario Clínico \\San Cecilio de Granada, Spain
    \And F. Herrera\\
  Andalusian Research Institute\\ in Data Science and\\ Computational Intelligence \\ University of Granada\\ 18071 
  Spain\\ 
}

\maketitle

\begin{abstract}

Currently, Coronavirus disease (COVID-19), one of the  most  infectious  diseases in the 21st century,  is  diagnosed using  RT-PCR testing,   CT scans and/or Chest X-Ray (CXR) images.  CT (Computed Tomography) scanners and RT-PCR testing are not available in most medical centers
and hence in many cases CXR images become the most time/cost effective tool for assisting clinicians in making  decisions. Deep learning neural networks have
a great potential for building COVID-19 triage systems and  detecting COVID-19 patients, especially patients with low severity. Unfortunately,  current databases do not allow building such systems as they are highly heterogeneous and  biased towards severe cases. 
This paper is three-fold: (i)  we demystify the high sensitivities achieved by most recent COVID-19 classification models, (ii) under a close collaboration with Hospital Universitario Clínico San Cecilio, Granada, Spain, we built COVIDGR-1.0, a homogeneous and balanced database that includes all levels of severity, from normal with Positive RT-PCR, Mild, Moderate to Severe. COVIDGR-1.0 contains 426 positive and 426 negative  PA (PosteroAnterior) CXR views and (iii) we propose COVID Smart Data based Network (COVID-SDNet)  methodology for improving the generalization capacity of COVID-classification models. Our approach reaches good and stable results with an accuracy of $97.72\% \pm 0.95 \%$, $86.90\% \pm 3.20\%$, $61.80\% \pm 5.49\%$ in severe, moderate and mild COVID-19 severity levels\footnote{Paper accepted for publication in Journal of Biomedical and Health Informatics}.  Our approach could help in the early detection of COVID-19.  COVIDGR-1.0 along with the severity level labels  are available to the scientific community through  this link \url{https://dasci.es/es/transferencia/open-data/covidgr/}.

\end{abstract}


%

\section{Introduction}

In the last months, the world has been witnessing how COVID-19 pandemic is increasingly infecting a large mass of people very fast everywhere in the world. The trends are not  clear yet but some research  confirm that this problem may persist until 2024 (\cite{kissler2020projecting}). Besides,  prevalence studies conducted in several countries reveal that a tiny  proportion of the population   have developed antibodies after exposure to the virus, e.g., 5\% in Spain \footnote{https://english.elpais.com/society/2020-05-14/antibody-study-shows-just-5-of-spaniards-have-contracted-the-coronavirus.html}. This means that frequently a large number of patients will need to be assessed in small time intervals by few number of clinicians and with very few  resources.

In general,    COVID-19  diagnosis is carried out using at least one of these three tests. 

\begin{itemize}
   
    \item   Computed Tomography
    (CT) scans-based assessment: it consists in analyzing 3D  radiographic images from different angles. The needed equipment for this assessment
    is not available in most hospitals and it takes more than 15 minutes per patient in addition to the time required for CT decontamination \footnote{//www.acr.org/Advocacy-and-Economics/ACR-Position-Statements/Recommendations-for-Chest-Radiography-and-CT-for-Suspected-COVID19-Infection}.
    
    \item Reverse Transcription Polymerase Chain Reaction (RT-PCR) test: it detects the viral RNA from sputum or nasopharyngeal swab (\cite{wong2020frequency}). It  requires specific material and equipment, which are not easily accessible and it takes at least 12 hours, which is not desirable as  positive COVID-19 patients should be identified and tracked as soon as possible.  Some studies  found that RT-PCR results from several tests at different points from the same patients were variable  during the course of the illness  producing a high false-negative rate  (\cite{li2020stability}). The authors suggested  that RT-PCR test should be combined with other clinical tests  such as CT.
    
    \item  Chest X-Ray (CXR):  The required equipment for this assessment are less  cumbersome and can be lightweight and transportable. In general, this  type of resources is more available than  the required for RT-PCR and CT-scan tests. In addition, CXR test takes about 15 seconds per patient (\cite{wong2020frequency}), which makes CXR  one of the most time/cost effective  assessment tools.
\end{itemize}

Few recent studies provide  estimates  on  expert  radiologists  sensitivity in the diagnosis of COVID-19 based on CT scans, RT-PCR and CXR.  A study on a set of 51 patients with chest CT and RT-PCR essay performed within 3 days, reported a  sensitivity in CT  of $98\%$ compared with RT-PCR sensitivity of $71\%$ (\cite{fang2020sensitivity}). A different study on 64 patients (26 men, mean age 56 $\pm$ 19 years)  reported a sensitivity of $69\%$  for CXR compared with $91\%$ for initial RT-PCR  (\cite{wong2020frequency}). According to an analysis of 636 ambulatory patients (\cite{weinstock2020chest}), most patients presenting to urgent care centers with confirmed coronavirus disease 2019 have normal or mildly abnormal findings on CXR. Only $58.3\%$ of these patients are correctly diagnosed by the expert eye. 

\begin{figure*}[h]
\centering
\includegraphics[width=0.6\textwidth] {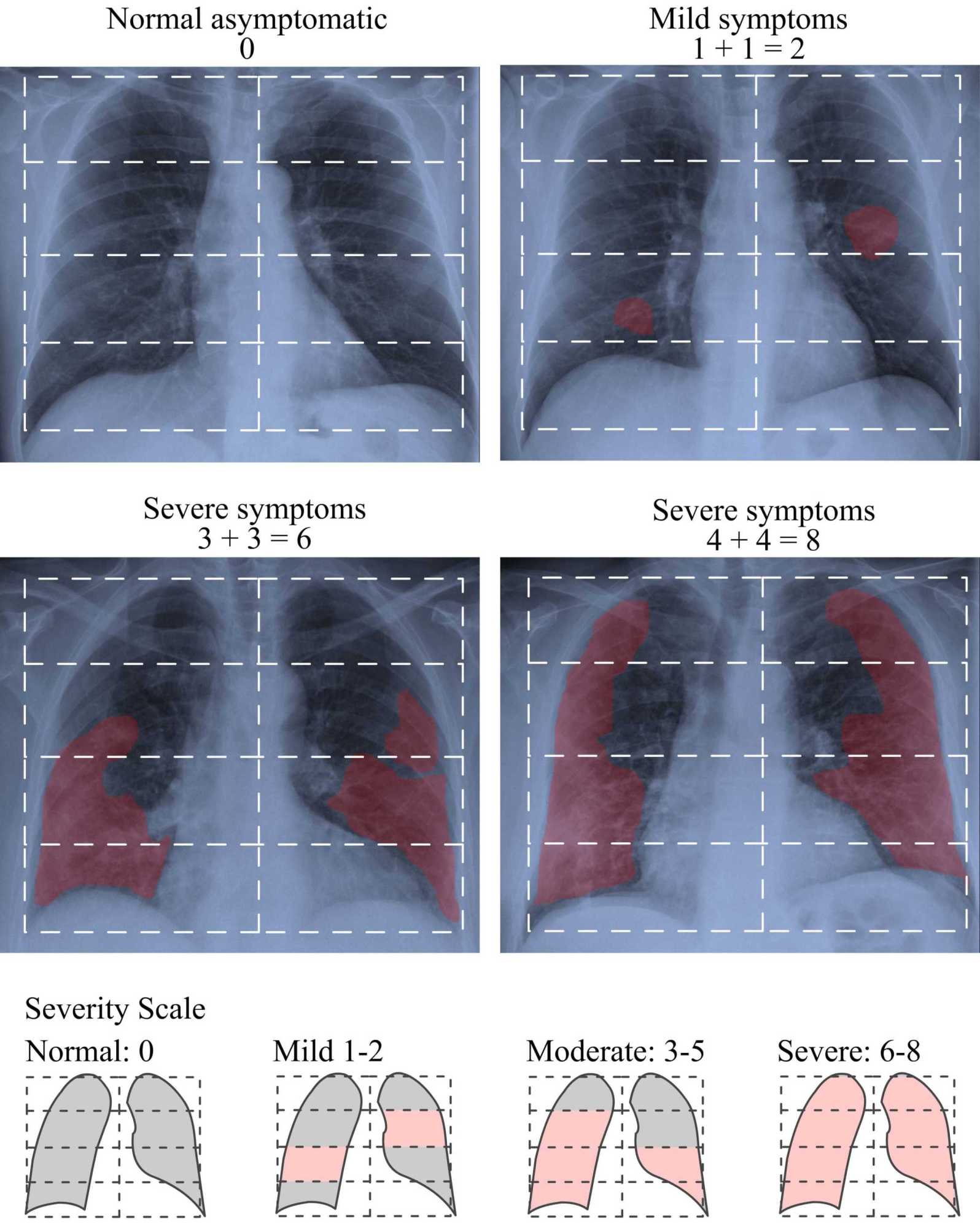}
\caption{The  stratification of radiological severity of COVID-19. Examples of how RALE index is calculated.}
\label{fig:strat}
\end{figure*}

In a recent study (\cite{wong2020frequency}), authors proposed  simplifying the quantification of the level of  severity  by adapting a  previously defined Radiographic Assessment of Lung Edema (RALE) score (\cite{warren2018severity}) to COVID-19. This new score is calculated by assigning a value between  0-4  to each lung depending on the extent of visual features such as, consolidation and ground glass opacities, in the four parts of each lung as  depicted in Fig. \ref{fig:strat}. Based on this score, experts can identify the level of severity of the infection among four severity stages, Normal 0, Mild 1-2, Moderate 3-5 and  Severe 6-8. In practice, a patient classified by expert radiologist as Normal  can have  positive RT-PCR. We refer to these cases as Normal-PCR+. Expert annotation adopted in this work is based in this score. 

Automated image analysis via Deep learning (DL) models have a great potential to optimize the role of CXR images for a fast diagnosis of COVID-19. A  robust and accurate DL model could serve as a triage method and as a support for medical decision making. An increasing number of recent works claim achieving impressive sensitivities  $>95\%$, far higher than  expert radiologists. These high sensitivities are due to the bias in the most used  COVID-19 dataset, {\it COVID-19  Image  Data  Collection} (\cite{cohen2020covid}). This dataset includes a very small number of COVID-19 positive cases, coming from highly heterogeneous sources (at least 15 countries) and most cases are severe patients,
an issue that drastically reduces its clinical value. 
To populate Non-COVID and Healthy classes, AI researchers are using CXR images from diverse pulmonary disease repositories. The obtained models will have no clinical value as well since they will be unable to detect patients with low and moderate severity, which are the target of a clinical triage system. In view of this situation, there is still a huge need  for higher quality datasets  built under the same clinical protocol and under a close collaboration with expert radiologists.

Multiple studies have proven that higher quality data ensures higher quality models. The concept of Smart Data refers to the process of converting raw data into higher quality data with higher concentration of useful information (\cite{LuengoGRGH20}). Smart data includes all pre-processing methods that improve value and  veracity of data. Examples of these methods include noise elimination,  data-augmentation (\cite{tabik2017snapshot}) and  data transformation (\cite{FuCiTNet20}) among other techniques.

In this work, we designed a  high clinical quality dataset, named COVIDGR-1.0 that includes  four levels of severity, Normal-PCR+, Mild, Moderate and Severe. We identified these four severity levels from a recent COVID-19 radiological  study (\cite{wong2020frequency}). We also propose COVID Smart Data based Network (COVID-SDNet) methodology. It combines segmentation, data-augmentation  and data transformations together with an appropriate Convolutional Neural Network (CNN) for inference.

The contributions of this paper can be summarized as follows:
\begin{itemize}
    \item We analyze reliability,  potential and limitations of the most used COVID-19 CXR datasets and models. 
    
    \item From a data perspective, we provide the first public dataset, called COVIDGR-1.0, that quantifies COVID-19 in terms of severity levels, normal, mild, moderate and severe, with the aim of building triage systems with high clinical value.
    
    \item From a pre-processing perspective, we combined several methods. To eliminate irrelevant information from the input CXR images, we used a new pre-processing method called segmentation-based cropping. To increase discrimination capacity of the classification model, we used a class-inherent transformation method inspired by GANs.
    
    \item From a post-processing perspective, we proposed a new inference process that fuses the predictions of the four transformed classes obtained by the class-inherent transformation method to calculate the final prediction.

    \item From a global perspective, we designed a novel methodology, named COVID-SDNet, with a high generalization capacity for COVID-19 classification based on CXR images. COVID-SDNet combines segmentation, data-transformation,  data-augmentation,  and a suitable CNN model together with an inference approach to get the final prediction.
    
\end{itemize}
 Experiments demonstrate that our  approach reaches good and stable results especially in moderate and severe levels,   with  $97.72\% \pm 0.95\%$ and $86.90\% \pm 3.20\%$ respectively.  Lower accuracies were obtained in mild and normal-PCR+ severity levels with $61.80\% \pm 5.49\%$ and $28.42\% \pm 2.58\%$, respectively.

This paper is organized as follows: A review of the most used datasets and   COVID-19 classification approaches is provided in Section \ref{sec:relatedWorks}.  Section {\color{black}\ref{sub:data}} describes how COVIDGR-1.0 is built and organized. Our approach is presented in Section {\color{black}\ref{sec:methodoloy}}. Experiments, comparisons and results are provided in Section {\color{black}\ref{sec:experimentsAndResults}}.   The inspection of  the model's decision using heatmaps is provided in Section \ref{sec:inspection} and the conclusions are pointed out in Section \ref{sec:conclusions}.

\section{Related works}
\label{sec:relatedWorks}
The last  months have known an increasing number of works exploring the potential of deep learning models for automating COVID-19  diagnosis based on  CXR images. The results are promising but still too much work needs to be done at the level of data and models design.  Given the potential bias in this type of problems, several studies include explication methods to their models. This section analyzes  the advantages and limitations of  current datasets an models for building automatic COVID-19 diagnosis systems with and without decision explication. 

\subsection{Datasets} 
There does not exist yet a high quality collection of CXR images for  building   COVID-19 diagnosis systems of high clinical value. Currently, the main source for  COVID-19 class is {\it COVID-19 Image Data Collection} (\cite{cohen2020covid}). It contains 76 positive and 26 negative PA views. These images were obtained  from highly heterogeneous equipment from all around the world. Another example  of COVID-19 dataset is Figure-1-COVID-19 Chest X-ray Dataset Initiative (\cite{agchung2020covid}). To build Non-COVID  classes, most studies are using CXR from one or multiple  public pulmonary disease data-sets.  Examples of these repositories are:
\begin{itemize}
    \item  RSNA Pneumonia CXR challenge dataset on Kaggle (\cite{2010nocovid}).
    \item ChestX-ray8 dataset  (\cite{wang2017chestx}).
    \item MIMIC-CXR dataset  (\cite{johnson2019mimic}).
    \item  PadChest dataset  (\cite{bustos2019padchest}).
\end{itemize}

For instance,  COVIDx 1.0 (\cite{wang2020covidnet}) was   built  by combining three public datasets: (i) {\it COVID-19 Image Data Collection} (\cite{cohen2020covid}), (ii) Figure-1-COVID- 19 Chest X-ray Dataset Initiative (\cite{agchung2020covid}) and (iii)   RSNA Pneumonia Detection Challenge dataset (\cite{2010nocovid}).   COVIDx 2.0 was built by  re-organizing  COVIDx 1.0 into three classes, Normal (healthy),  Pneumonia and COVID-19, using 201 CXR images for COVID class, including  PA(PosteroAnterior) and AP(AnteroPosterior) views (see  Table \ref{Fig:COVIDx}). Notice  that for a correct learning front view (PA)  and back view (AP)  cannot be mixed in the same class.

\begin{table}[h]
\begin{tabular}{|l||l|l|l|}
\hline

Version & Normal(healthy) & Pneumonia & COVID-19 \\\hline\hline
1.0 & 1,583 & 4,273 (Bacterial+viral) & 76\\\hline
2.0 &8,066 & 8,614 & 190\\\hline
\end{tabular}
\caption{ A brief description of COVIDx dataset (\cite{cohen2020covid}) (only PA views are counted).}
\label{Fig:COVIDx}
\end{table}

Although the value of  these datasets is unquestionable as they are being useful for carrying out first studies and  reformulations, they do not guarantee useful triage systems for the next reasons. It is not clear what annotation protocol has been followed for constructing the positive class in {\it COVID-19 Image Data Collection}. The included data is highly heterogeneous and hence DL-models can rely on other aspects than COVID visual features to differentiate between the involved classes. This dataset  does not provide a representative spectrum of COVID-19 severity levels, most positive cases  are of  severe patients (\cite{kundu2020might}).   In addition, an interesting critical analysis of these datasets has shown that CNN models obtain similar results  with and without eliminating most of the lungs in the input X-Ray images \cite{maguolo2020critic}, which  confirms again that there is a huge need of COVID-19 datasets with high clinical value.

Our claim is that the design of a high quality dataset must be done under a close collaboration between expert radiologists and AI experts. The annotations must follow the same  protocol and  representative numbers of all levels of severity, especially  Mild and Moderate levels, must be included.



\begin{table*}[!h]
\resizebox{\textwidth}{!}{%
\begin{tabular}{|l||p{3.5cm}|l|p{2.5cm}|l|l|l|}
\hline
 Ref.              & Classes              & Datasets     & Model    & Partition  & Sens. & Acc.   \\\hline\hline
 (\cite{wang2020covidnet})& Normal, Pneumonia, COVID & COVIDx 1.0& COVIDNet & 98\% - 2\%  & 87.1\% & 92.6\% \\\hline

(\cite{afshar2020covid}) &Normal, COVID & COVIDx 1.0& COVID-CAPS& 98\% - 2\%  &  90\% & 95.7\% \\\hline

\multirow{2}*{(\cite{ozturk2020automated})}  & No-Findings, COVID & \multirow{2}*{(\cite{cohen2020covid}) + (\cite{wang2017chestx})}  &   \multirow{2}*{DarkCovidNet} & 5-FCV  &  90.65\% & 98.08\%    \\
                            & No-Findings, Pneumonia, COVID  &    & & 5-FCV & 97.9\% & 87.02\%  \\ \hline

(\cite{karim2020deepcovidexplainer})  & Normal, Pneumonia, COVID & COVIDx 2.0+(\cite{2010nocovid}) & VGG-19 + DenseNet-161  & 70\% - 30\%  & 93\% & 96.77\% \\\hline
(\cite{ghoshal2020estimating})&Normal, Bacterial, Viral, COVID & (\cite{cohen2020covid})+(\cite{2010nocovid})& Bayesian ResNet50V2&80\% - 20\% & 85.71\% & 89.82\% \\\hline
(\cite{apostolopoulos2020covid}) & Normal, Pneumonia, COVID & \cite{cohen2020covid} + (\cite{2010nocovid})+ other sources & MobileNet & 10-FCV & 98.66\% & 96.78\% \\\hline
\end{tabular}}
 \caption{Summary of related works that analyze variations of COVIDx with CNN. }
\label{Fig:RW}
\end{table*}

\begin{table*}[b]
\centering
\begin{tabular}{|l||l||l|l|l|l|}
\hline
Dataset   & Class & \#images   & women & men & \#img. per severity level \\\hline\hline
COVIDGR-1.0  & Negative &  426 & 239& 187& \\\cline{2-6}
          & COVID-19 & 426  &190 & 236 & Normal-PCR+: 76\\
         &  &   &  &   & Mild: 100\\
         &  &   &  &   & Moderate: 171\\
         &  &   &  &   & Severe: 79\\\hline
\end{tabular}
 \caption{A brief summary of COVIDGR-1.0 dataset. All samples in COVIDGR 1.0 are segmented   CXR images considering only PA view. }
\label{Fig:COVIDGR}
\end{table*}

\subsection{DL classification models} 
Existing  related works are not directly comparable as  they consider different combinations of public data-sets and different experimental setup. A  brief summary of these works is provided  in Table \ref{Fig:RW}.

The most related studies to ours as they proposed different models to the typical ones are (\cite{wang2020covidnet}) and  (\cite{afshar2020covid}). In (\cite{wang2020covidnet}), the authors designed a deep network, called COVIDNet. 
They affirmed that COVIDNet reaches an overall accuracy of 92.6\%, with  97.0\% sensitivity in Normal class, 90.0\% in Non-COVID-19 and 87.1\% in COVID-19. The authors of a smaller network, called  COVID-CAPS (\cite{afshar2020covid}), also claim that their model achieved an accuracy of $98.7\%$, sensitivity of $90\%$,  and specificity of $95.8\%$. These results look too impressive when compared to expert radiologist sensitivity,  $69\%$. This can be explained by the fact that the used dataset is biased to severe COVID cases  (\cite{kundu2020might}). In addition,  the performed experiments in both cited works are not statistically reliable as they were evaluated on one single partition. The stability of these models, in terms of standard deviation,  has not been reported.

\subsection{DL classification models with explanation approaches} 
Several interesting explanations were proposed to help inspect the predictions  of DL-models (\cite{ghoshal2020estimating,karim2020deepcovidexplainer}) although all  their classification models were  trained and validated  on  variations of COVIDx. The authors in  (\cite{karim2020deepcovidexplainer}) first use an ensemble of two CNN networks to predict the class of the input image, as Normal, Pneumonia or COVID. Then  highlight class-discriminating regions in the input CXR image using gradient-guided class activation maps (Grad-CAM++) and layer-wise relevance propagation (LRP).
In (\cite{ghoshal2020estimating}), the authors proposed explaining the decision of the classification model to radiologists using different saliency map types together with uncertainty estimations (i.e., how certain is the model in the  prediction).

\section{COVIDGR-1.0: Data acquisition, annotation and organization}
\label{sub:data}

Instead of starting with an extremely large and noisy dataset, one can  build a small and smart dataset then augment it in a way it increases the performance of the model. This approach has proven effective in multiple studies. This is particularly true in the medical field, where access to data is heavily protected due to privacy concerns and costly expert annotation.

 Under a close  collaboration with  four highly trained  radiologists from  Hospital Universitario Clínico San Cecilio, Granada, Spain, we first established a protocol on how CXR images are selected and annotated to be included  in the dataset.  A CXR image is annotated as COVID-19 positive if both  RT-PCR test and expert radiologist confirm that decision within less than 24 hours. CXR with positive PCR that were annotated by expert radiologists as Normal are labeled as Normal-PCR+. The involved radiologists  annotated 
 the level of severity of positive cases based on RALE score as: Normal-PCR+, Mild, Moderate and Severe. 

 COVIDGR-1.0 is organized into two classes, positive and negative.  It contains   852 images distributed into 426 positive  and 426 negative cases, more details are provided in Table \ref{Fig:COVIDGR}.  All the images were obtained from the same equipment and under the same X-ray regime.  Only PosteriorAnterior (PA) view is considered. COVIDGR-1.0  along with the severity level labels are available to the scientific community through  this  link: \url{
 https://dasci.es/es/transferencia/open-data/covidgr/}.

\begin{figure*}[t]
\centerline{\includegraphics[width=0.40\textwidth] {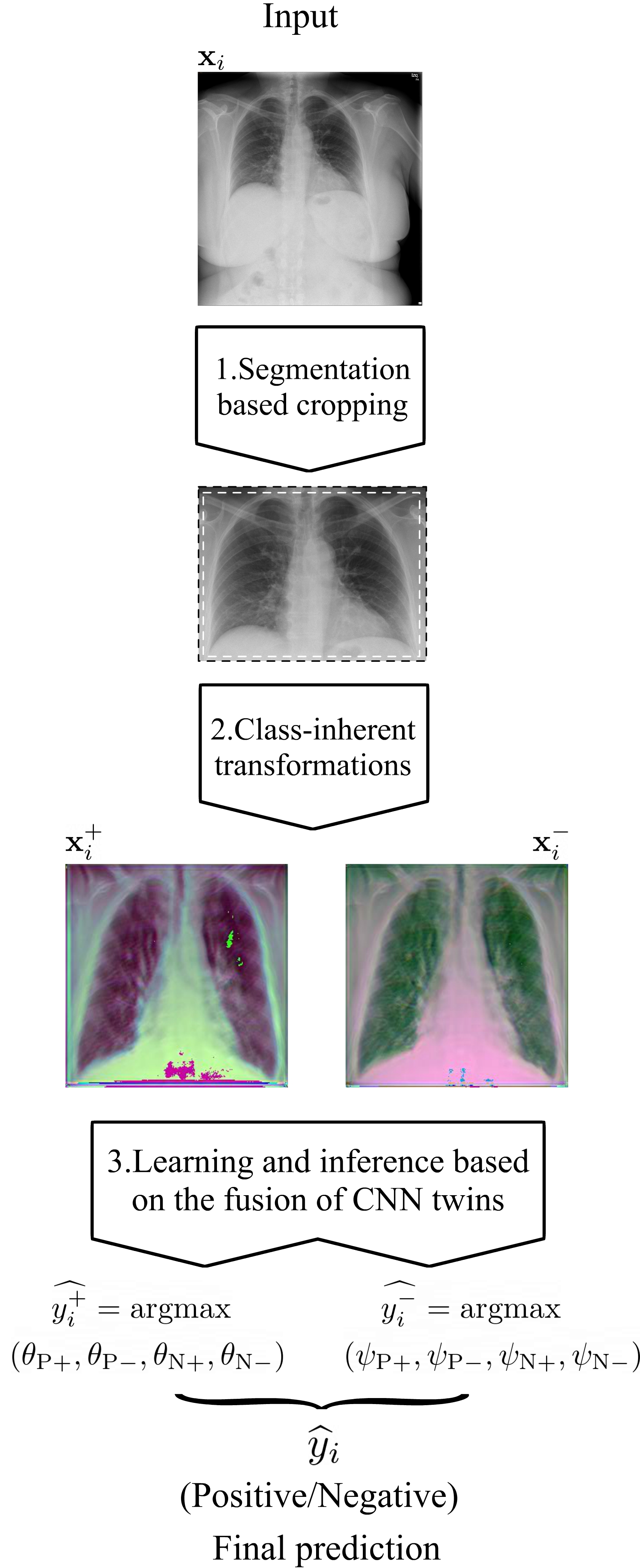}}
\caption{\color{black} Flowchart of the proposed COVID-SDNet methodology.}
\label{fig:workflow}
\end{figure*}    
\section{COVID-SDNet methodology}
\label{sec:methodoloy}

In this section, we describe  COVID-SDNet  methodology  in detail, covering  pre-processing to produce smart data, including segmentation and data transformation for increasing discrimination between positive and negative classes, combined with a deep CNN for classification.

One of the pieces of COVID-SDNet is the CNN-based classifier. We have selected Resnet-50 initialized with ImageNet weights for a transfer learning approach. To adapt this CNN to our problem, we have removed the last layer of the net and added a 512 neurons layer with ReLU activation and a two or four neurons layer (according to the considered number of classes) with softmax activation. 

 Let $X$ be the set of $n$ images  and $K$ the total number of classes. 
Each image $\mathbf{x}_i \in X$ has a true label $y_i$ with $i = 1,2,\dots,n$.  The softmax function computes the probability that an image belongs to class $k$ with $k = 1,\dots,K$. Let $\mathbf{w} = (w_1,\dots,w_K)$ be the output of the last fully connected layer before the softmax activation is applied. Then, this function is defined as: $\mathrm{softmax}: \mathbb{R}^K \rightarrow [0,1]^K$,
\begin{equation*}
    \mathrm{softmax}(\mathbf{w})_j = \frac{\mathrm{exp}(w_j)}{\sum_{k=1}^K \mathrm{exp}(w_k)} \enspace .
\end{equation*}
Let $\widehat{y_i}$ be the class prediction of the network for the image $\mathbf{x}_i$, then $\widehat{y_i} = \mathrm{argmax}(\mathrm{softmax}(\mathbf{w}))$, where $\mathbf{w}$ is the output vector of the last layer before softmax is applied for the input $\mathbf{x}_i$.

All the layers of the network were fine-tuned. We used a batch size of 16 and SGD as optimizer. 


The main stages of COVID-SDNet are three, two  associated to pre-processing for producing  quality data (smart data stages) and the  learning and inference process. A flowchart of  COVID-SDNet is depicted in      Fig.  \ref{fig:workflow}.


\begin{enumerate}

\item {\it Segmentation-based cropping: Unnecessary information elimination }

\begin{figure*}[t]
\centering
\subfloat[Input image]{\includegraphics[ width=0.27\textwidth] {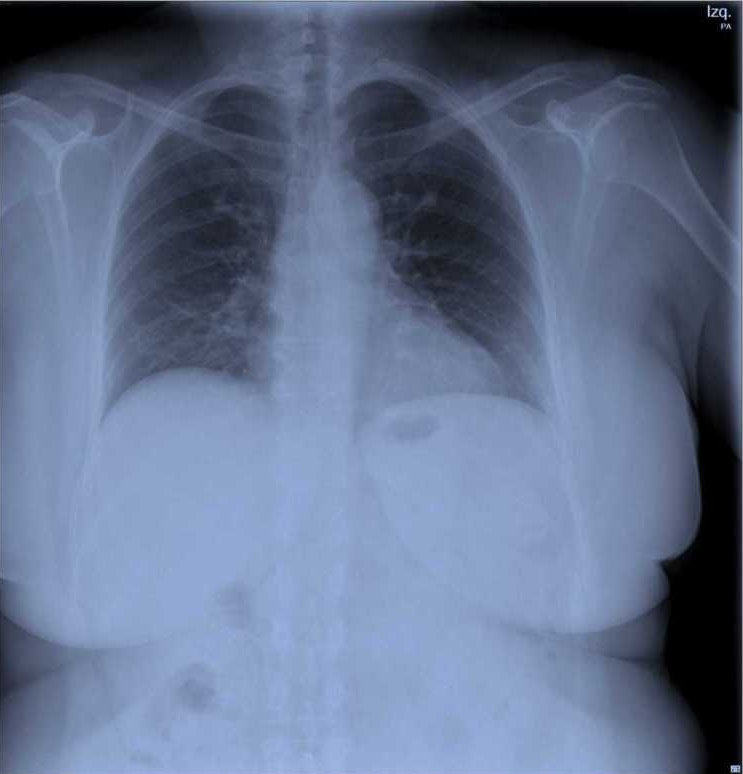}}
\hspace{0.1cm}
\subfloat[The smallest rectangle that delimits the left and right segmented lungs is calculated]{\includegraphics[ width=0.27\textwidth] {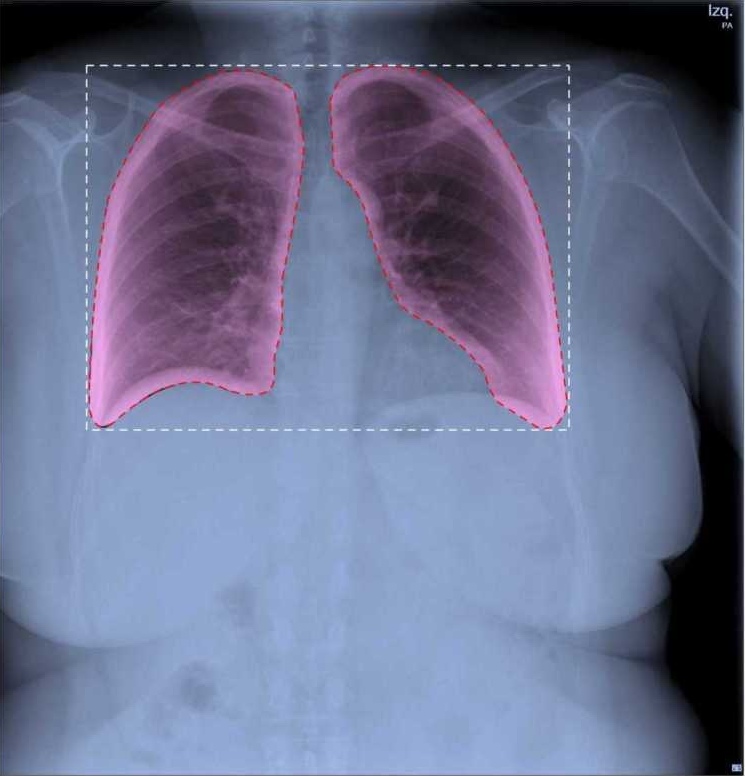}}
\hspace{0.1cm}
\subfloat[2.5\% of pixels are added to the left, right, up and down sides of the rectangle then  the final rectangle is cropped]{\includegraphics[ width=0.27\textwidth] {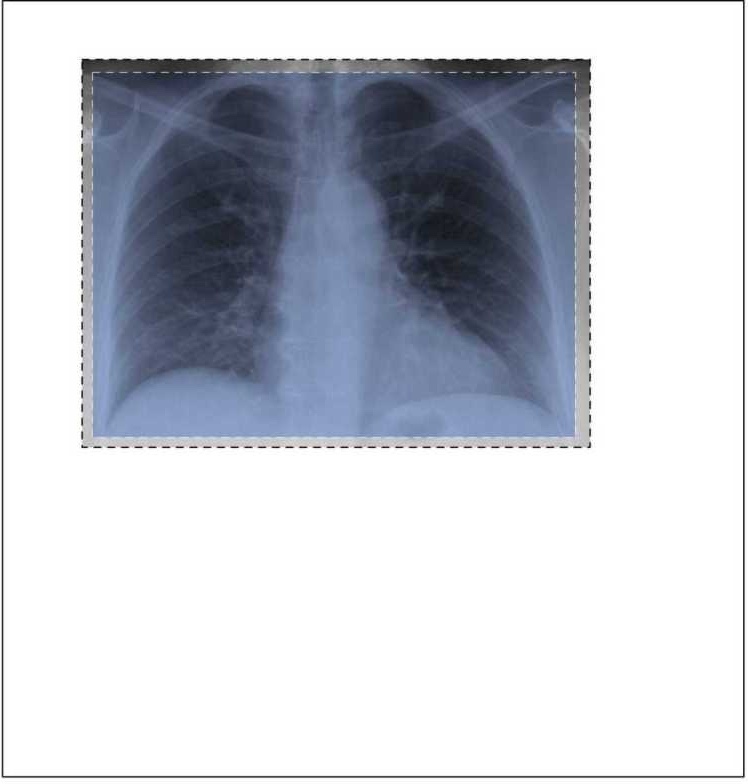}}
\caption{The segmentation-based cropping pre-processing applied to the input X-ray image}
\label{fig:seg}
\end{figure*}


Different CXR equipment brands include different extra information about the patient  in the sides and contour of CXR images. The position and size of the patient may also imply the inclusion of more parts of the body, e.g., arms, neck, stomach. As this information may alter the learning of the classification model, first, we segment the lungs using
the  U-Net segmentation model provided in (\cite{Kaggle-seg}),   pre-trained on Tuberculosis Chest X-ray Image datasets (\cite{jaeger2014two}) and  RSNA Pneumonia CXR challenge dataset (\cite{2010nocovid}). Then, we calculate the smallest rectangle that delimits the left and right segmented-lungs. Finally,  to avoid eliminating useful information, we add  $2.5\%$ of pixels to the left, right, up and down sides of the rectangle. The resulting rectangle is cropped. An illustration with example of this pre-processing  is shown in Fig. \ref{fig:seg}.

\item {\it Class-inherent transformations Network}\\

\begin{figure*}[t]
\centering
\subfloat[Original Negative]{\includegraphics[ width=0.27\textwidth] {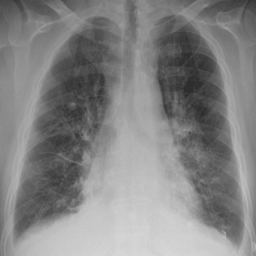}}
\hspace{0.1cm}
\subfloat[Negative transf.]{\includegraphics[ width=0.27\textwidth] {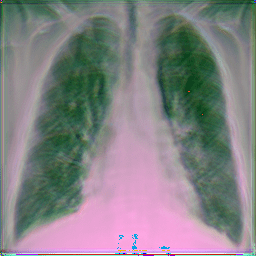}}
\hspace{0.1cm}
\subfloat[Positive transf.]{\includegraphics[ width=0.27\textwidth] {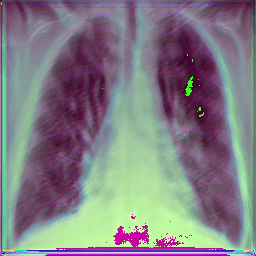}}
\caption{Class-inherent transformations applied to a negative sample. a) Original negative sample; b) Negative transformation; c) Positive transformation}
\label{fig:CIT}
\end{figure*}

 To increase the discrimination capacity of the classification model,  we used, {\color{black} FuCiTNet (\cite{FuCiTNet20})},  a Class-inherent transformations (CiT) Network  inspired by GANs (Generative Adversarial Networks). {\color{black}This transformation method is actually an array of two generators $G_\mathrm{P}$ and $G_\mathrm{N}$, where $_\mathrm{P}$ refers to the positive class and $_\mathrm{N}$ refers to the negative class.  $G_\mathrm{P}$ learns the inherent-class transformations of the positive  class $\mathrm{P}$ and $G_\mathrm{N}$ learns the inherent-class transformations of the negative class $\mathrm{N}$. In other words,  $G_\mathrm{P}$ learns the transformations that bring an input image from its own  $k$ domain, with $k \in \{\mathrm{P},\mathrm{N}\}$,  to the  $\mathrm{P}$ class domain. Similarly, $G_\mathrm{N}$ learns the transformations that bring the input image from its $k$ space, with $k \in \{\mathrm{P},\mathrm{N}\}$,  to the $\mathrm{N}$ class space.} The classification loss is introduced in the generators to drive the learning of each specific $k$-class transformations. {\color{black} 
 That is, each generator is optimized based on the following  loss function:
  \begin{equation}
   \mathcal{L}_{gen_k} =  l_{MSE} + 0.006\cdot l_{Perceptual} + \lambda\cdot l_{CE}(\mathrm{y}==k)
\label{eq:system_loss}
\end{equation}

Where $l_{MSE}$ is a pixel-wise Mean Square Error, $l_{Perceptual}$ is a perception Mean Square Error and $l_{CE}$ is the classifier loss. The weighted factor $\lambda$  indicates how much the generator must change its outcome to suit the classifier.} More details about these transformation networks can be found in  (\cite{FuCiTNet20}).

The  architecture  of  the  generators   consists  of 5 identical residual blocks.  Each block has two convolutional layers with $3\times3$ kernels and $64$ feature maps followed by batch-normalization layers and Parametric ReLU as activation function. The last residual block is followed by a final convolutional layer which reduces the output image channels to 3 to match the input’s dimensions. The classifier is a ResNet-18 which consists of an initial convolutional layer with  $7\times7$  kernels  and  $64$  feature  maps  followed  by  a  $3\times3$  max  pool  layer. Then, 4 blocks of two convolutional layers with $3\times 3$ kernels with 64, 128, 256 and 512 feature maps respectively followed by a $7\times 7$ average pooling and one fully connected layer which outputs a vector of $N$ elements.  ReLU is used as  activation function.

{\color{black} Once the generators learn the corresponding transformations, the dataset is processed using $G_\mathrm{P}$ and $G_\mathrm{N}$. Two pair of images $(\mathbf{x}_i^+, \mathbf{x}_i^-)$ will be obtained from each input image $\mathbf{x}_i$, $i=1,\dots,n$, where  $\mathbf{x}_i^+$ and $\mathbf{x}_i^-$ are respectively the positively and negatively transformed images of $\mathbf{x}_i$. Note that, once the entire dataset is processed, we have four classes ($\mathrm{P}+, \mathrm{P}-, \mathrm{N}+, \mathrm{N}-$) instead the original $\mathrm{P}$ and $\mathrm{N}$ classes. Let $y_i$ be the class of $\mathbf{x}_i$, $y_i \in \{\mathrm{P}, \mathrm{N}\}$. If $y_i=\mathrm{P}$, $G_\mathrm{P}$ and $G_\mathrm{N}$  will produce  the positive transformation $\mathbf{x}_i^+$ with $y_i^+=\mathrm{P+}$ and the negative transformation $\mathbf{x}_i^-$ with $y_i^-=\mathrm{P-}$, respectively. If $y_i=\mathrm{N}$, $G_\mathrm{P}$ and $G_\mathrm{N}$ will produce the positive transformation $\mathbf{x}_i^+$ with $y_i^+=\mathrm{N+}$ and the negative transformation $\mathbf{x}_i^-$ with $y_i^-=\mathrm{N-}$, respectively. Fig.  \ref{fig:CIT} illustrates with example the transformations applied by $G_\mathrm{N}$ and $G_\mathrm{P}$.} Notice that these transformations are not meant to be interpretable by the human eye but rather  help the classification model  better distinguish between the different classes.


\item {\it Learning and inference based on the fusion of CNN twins}\\

The CNN classification model described above in this section (Resnet-50) is trained to predict the new four classes: $\mathrm{P}+, \mathrm{P}-, \mathrm{N}+, \mathrm{N}-$.  {\color{black} The output of the network (after softmax is applied) for each transformed image associated to the original one is a vector $\boldsymbol{\theta}=(\theta_{\mathrm{P+}}, \theta_{\mathrm{P-}}, \theta_{\mathrm{N+}}, \theta_{\mathrm{N-}})$, where $\theta_j$ is the probability of the transformed image to belong to class $j \in \{\mathrm{P+}, \mathrm{P-}, \mathrm{N+}, \mathrm{N-}\}$. Herein, we propose an inference process to fuse the output of the two transformed images $\mathbf{x}_i^+$ and $\mathbf{x}_i^-$ to predict the label of the original image $\mathbf{x}_i$. In this way, for each pair  $(\mathbf{x}_i^+, \mathbf{x}_i^-)$, the prediction of the original image $\widehat{y_i}$ will be either $\mathrm{P}$ or $\mathrm{N}$. Let $\widehat{y_i^+} = \mathrm{arg max}~ \boldsymbol{\theta} = \mathrm{arg max}~ \mathrm{(\theta_{\mathrm{P+}}, \theta_{\mathrm{P-}}, \theta_{\mathrm{N+}}, \theta_{\mathrm{N-}})}$ and $\widehat{y_i^-} = \mathrm{arg max}~ \boldsymbol{\psi} = \mathrm{arg max}~ \mathrm{(\psi_{\mathrm{P+}}, \psi_{\mathrm{P-}}, \psi_{\mathrm{N+}}, \psi_{\mathrm{N-}})}$ be the ResNet-50 predictions for $\mathrm{x}_i^+$ and $\mathrm{x}_i^-$ respectively. Then:
  
    \begin{enumerate}
    	\item If $\widehat{y_i^+}$ = $\mathrm{N+}$ and $\widehat{y_i^-}$ = $\mathrm{N-}$, then $\widehat{y_i}$ = N.
    	\item If $\widehat{y_i^+}$ = $\mathrm{P+}$ and $\widehat{y_i^-}$ = $\mathrm{P-}$, then $\widehat{y_i}$ = P.
    	\item If none of the above applies, then  
    	\begin{equation*}
    	\widehat{y_i} = \left\lbrace 
                    \begin{array}{ll}
                      \mathrm{N}~ \mathrm{if}~ \mathrm{max} (\theta_{\mathrm{N}j}, \psi_{\mathrm{N}j}) > \mathrm{max}(\theta_{\mathrm{P}j}, \psi_{\mathrm{P}j}),\\   \enspace \enspace j \in \{+ ,  -\} \\
                      \mathrm{P}~ \mathrm{otherwise} \enspace .
                    \end{array}	
    	\right. 
    	\end{equation*}
    \end{enumerate}}
    Experimentally, we used a batch size of 16 and SGD as optimizer. 

\end{enumerate}

\section{Experiments and Results}
\label{sec:experimentsAndResults}
In this section we (1) provide all the information about the used experimental setup, (2)   evaluate two state-of-the-art  COVID classification models and {\color{black} FuCiTNet alone (\cite{FuCiTNet20})}  on our dataset then,  analyze (3) the impact of data pre-processing  and  (4) Normal-PCR+ severity level on our approach.

\subsection{Experimental setup}

Due to the high variations between different executions, we performed   5 different 5 fold cross validations in all the experiments. Each experiment uses 80\% of  COVIDGR-1.0 for training and the remaining 20\% for testing. To choose when to stop the training process, we used a random 10\% of each training set for validation. In each experiment, a proper set of data-augmentation techniques is carefully selected. All results, in terms of sensitivity, specificity, precision, F1 and accuracy, are presented using the average values and the standard deviation of the 25 executions.  The used metrics are calculated as follows:

\[\operatorname{recall}(\text{positive class}) = \mathit{sensitivity}=\frac{\text{TP}}{\text{actual positives}}\]

\[\operatorname{recall}(\text{negative class}) = \mathit{specificity} = \frac{\text{TN}}{\text{actual negatives}}\]

\[\operatorname{precision}(\text{positive class}) = \frac{\text{TP}}{\text{predicted positives}}\] 

\[\operatorname{precision}(\text{negative class}) = \frac{\text{TN}}{\text{predicted negatives}}\] 

\[\mathit{accuracy} = \frac{\text{ TP+TN}}{\text{total predictions}}\]

\[\operatorname{F1}= 2 \cdot \frac{\text{precision $\cdot$ recall}}{\text{precision + recall}} \]

 TP and TN refers respectively to the number of true positives and true negatives.

\subsection{Analysis of COVIDNet and COVID-CAPS}

\begin{table*}[t]
\centering
\begin{tabular}{|l|c|c|c|c|c|}
\hline
Class                        & \multicolumn{2}{c|}{Negative} & \multicolumn{2}{c|}{Positive (COVID-19)} & \multicolumn{1}{l|}{\multirow{2}{*}{Accuracy}} \\ \cline{1-5}
Metric                       & Specificity      & Precision     & Sensitivity       & Precision      & \multicolumn{1}{l|}{}                          \\ \hline
COVIDNet-CXR A (\cite{wang2020covidnet}) & 0.23 &	16.00 &	\bfseries{99.29} &	33.54 &	49.76 \\ \hline
Retrained COVIDNet-CXR A & \bfseries{88.82$\pm$0.90}&	3.36$\pm$6.15&	46.82$\pm$17.59&	\bfseries{81.65$\pm$6.02}&		\bfseries{67.82$\pm$6.11} \\ \hline
COVID-CAPS  (\cite{afshar2020covid})  & 26.30&	45.81&		69.01&	48.36&		47.66    \\ \hline
Retrained COVID-CAPS & 65.74$\pm$9.93&	\bfseries{65.62$\pm$3.98}&		64.93$\pm$9.71&	66.07$\pm$4.49&		65.34$\pm$3.26 \\ \hline
\end{tabular}
\caption{COVIDNet and COVID-CAPS results on our dataset}
\label{tab:CompOurs}
\end{table*}

\begin{table*}[t]
\resizebox{\textwidth}{!}{\begin{tabular}{|c|c|c|c|c|c|c|c|}
\hline
Class    & \multicolumn{3}{c|}{N} & \multicolumn{3}{c|}{P} & \multirow{2}{*}{Accuracy} \\ \cline{1-7}
Metric     & {\bf Specificity}         & Precision      & F1             & {\bf Sensitivity}         &  Precision      & F1             &                   \\ \hline
COVIDNet-CXR  & 88.82$\pm$0.90&	3.36$\pm$6.15&73.31$\pm$3.79&	46.82$\pm$17.59&	81.65$\pm$6.02&56.94$\pm$15.05&		67.82$\pm$6.11\\ \hline
COVID-CAPS &  65.74$\pm$9.93&	65.62$\pm$3.98&65.15$\pm$5.02&		64.93$\pm$9.71&	66.07$\pm$4.49& 64.87$\pm$4.92&		65.34$\pm$3.26 \\ \hline\hline
Without seg. & 79.87$\pm$8.91 & 71.91$\pm$3.12 & 75.40$\pm$4.91 & 68.63$\pm$6.08 & 78.75$\pm$6.31 & 72.689$\pm$3.45 & 74.25$\pm$3.61\\ \hline
With seg.  &  78.41$\pm$7.09  & 73.36$\pm$4.66 & 75.46$\pm$2.97 & 70.80$\pm$8.26 & 77.17$\pm$4.79 & 73.40$\pm$4.01 & 74.60$\pm$2.93 \\ \hline
FuCiTNet &  \textbf{80.79$\pm$6.98} & 72.00$\pm$4.48 & 75.84$\pm$3.18 & 67.90$\pm$8.58 & 78.48$\pm$4.99 & 72.35$\pm$4.76 & 74.35$\pm$3.34\\ \hline
\begin{tabular}[c]{@{}c@{}}COVID-SDNet\end{tabular} 
&   79.76$\pm$6.19  & {\bf 74.74$\pm$3.89} & {\bf 76.94$\pm$2.82} & {\bf72.59$\pm$6.77} & {\bf 78.67$\pm$4.70} & {\bf 75.71$\pm$3.35} & {\bf 76.18$\pm$2.70} \\ \hline
\end{tabular}}
\caption{Results of COVID-19 prediction using Retrained COVIDNet-CXR A, Retrained COVID-CAPS,  ResNet-50 with and without segmentation,  FuCiTNet and COVID-SDNet. All  four levels of severity in the positive class are taken into account.}
\label{tab:binary}
\end{table*}

We compare our approach with the two most related approaches to ours, COVIDNet (\cite{wang2020covidnet}) and COVID-CAPS (\cite{afshar2020covid}). 
\begin{itemize}
    \item COVIDNet: Currently, the authors of this network  provide three versions, namely A, B and C, available at (\cite{COVIDNet}). A has the largest number of trainable parameters, followed  by B and C. We performed two evaluations of each network  in such a way that the results will be comparable to ours. 
    \begin{itemize}
        \item First, we tested COVIDNet-A, COVIDNet-B and COVIDNet-C, pre-trained on COVIDx, directly on our dataset by considering only  two classes: Normal (negative), and COVID-19 (positive). The whole dataset (426 positive images and 426 negative images) is evaluated. We report in 
        \autoref{tab:CompOurs} recall and precision results for Normal and COVID-19 classes.

        \item Second, we retrained COVIDNet on our dataset. It is important to note that as only a checkpoint of each  model is available, we could not remove the last layer of these networks, which has three neurons. We used 5 different 5 fold cross validations. In order to be able to retrain  COVIDNet models, we had  to add a third Pneumonia class into our dataset.  We randomly selected  426 images from the Pneumonia class in COVIDx dataset. We used the same hyper-parameters as the ones indicated in their training script, that is, 10 epochs, a batch size of 8 and a learning rate of 0.0002. We changed covid\_weight to 1 and covid\_percent to 0.33 since we had the same number of images in all the classes. Similarly,  we report in \autoref{tab:CompOurs} recall and precision of our two classes, Normal and COVID-19, and omit recall and precision of Pneumonia class. The accuracy reported in the same table only takes into account the images from our two classes. As with our models, we report here the mean and standard deviation of all metrics.
    \end{itemize}
Although we analyzed all three A, B and C variations of COVIDNet, for simplicity we only report the results of the best one.

    \item COVID-CAPS: This is a capsule network-based model proposed in (\cite{afshar2020covid}). Its architecture is notably smaller than COVIDNet, which implies a dramatically lower number of trainable parameters. Since the authors also provide a checkpoint with weights trained in the COVIDx dataset, we were able to follow a similar procedure than with COVIDNet:
  
    \begin{itemize}
        \item First, we tested the  pretrained weights using COVIDx on  COVIDGR-1.0 dataset. COVID-CAPS is designed to predict two classes, so we reused the same architecture with the new dataset and compute the evaluation metrics shown in \autoref{tab:CompOurs}.
        
        \item Second, COVID-CAPS architecture was retrained over the COVIDGR-1.0 dataset. This process finetunes the weights to improve class separation. The retraining process is performed using the same setup and hyper-parameters reported by the authors.  Adam optimizer is used across 100 epochs with a batch size of 16. Class weights were omitted as with COVIDNet, since this dataset contains balanced classes in training as well as in test. Evaluation metrics are computed for five sets of 5-fold cross-validation test subsets and summarized in \autoref{tab:CompOurs}.
    \end{itemize}
\end{itemize}

The results from  Table~\ref{tab:CompOurs} show that COVIDNet and COVID-CAPS trained on COVIDx overestimate COVID-19 class in our dataset, i.e., most images are classified as positive, resulting in very high sensitivities but at the cost of low positive predictive value. However, when COVIDNet and COVID-CAPS are re-trained on COVIDGR-1.0 they achieve slightly better overall accuracy and a higher balance between sensitivity and specificity, although they seem to acquire a bias favoring the negative class. In general, none of these models perform adequately for the detection of the disease from CXR images in our dataset.

\subsection{Results and Analysis of COVID prediction}

The results of  the baseline COVID classification model considering all the levels of severity,    with and without segmentation,  FuCiTNet (\cite{FuCiTNet20}), and COVID-SDNet are shown in Table \ref{tab:binary}.

In general, COVID-SDNet achieves better and more stable results than the rest of approaches. In particular, COVID-SDNet achieved the highest balance between specificity and sensitivity with $76.94 \pm 2.82$ F1 in the negative class and $75.71 \pm 3.35$ F1 in the positive class. Most importantly, COVID-SDNet   achieved the best 
sensitivity $72.59\pm 6.77$  and accuracy with $76.18\pm 2.70$. FuCiTNet provides in general good  but lower and less stable results than COVID-SDNet. When comparing the results of the baseline classification model with and without segmentation, we can observe that the use of segmentation improves substantially the sensitivity, which is the most important criteria for a triage system. This can be explained by the fact  that segmentation allows the model to focus on  most important parts of the CXR image.

\subsection*{Analysis per  severity level}

To determine which levels are the hardest to distinguish by the best approach, we have analyzed the  accuracy per severity level (S), with $\mathrm{accuracy (S)} = \frac{\mathrm{Correct ~predictions(S)} }{\mathrm{Total ~number(S)}} \enspace$, where  $S \in \{$Normal-PCR+, Mild, Moderate, Severe$\}$.  The results are shown in Table \ref{tab:levels}.
 
\begin{table}[h]
\centering
\begin{tabular}{|l|c|}
\hline
S (Severity level)      &  accuracy (S)($\%$)  \\\hline \hline 
  Normal-PCR+ & 28.42 $\pm$ 2.58 \\
  Mild & 61.80 $\pm$ 5.49 \\
  Moderate & 86.90 $\pm$ 3.20 \\
  Severe & 97.72 $\pm$ 0.95 \\\hline
\end{tabular}
\caption{Results of  COVID-SDNet per severity level.}
\label{tab:levels}
\end{table}

As it can be seen from these results, COVID-SDNet correctly distinguish Moderate and Severe levels with  an accuracy of $86.90\%$ and $97.72 \%$, respectively. This is due to the  fact that 
Moderate and Severe CRX images contain more important visual features than Mild and Normal-PCR+ which ease the classification task. Normal-PCR+ and Mild cases are much more difficult to identify as they contain few or none visual features. These results are coherent with  the clinical studies provided in (\cite{weinstock2020chest}) and (\cite{wong2020frequency}) which report that expert sensitivity is very low in Normal-PCR+ and Mild infection levels. Recall that the expert eye does not see any visual signs  in  Normal-PCR+  although the PCR is positive. Those cases are actually considered as asymptomatic patients. 

\begin{table*}[t]
\resizebox{\textwidth}{!}{\begin{tabular}{|c|c|c|c|c|c|c|c|}
\hline
Class    & \multicolumn{3}{c|}{N} & \multicolumn{3}{c|}{P} & \multirow{2}{*}{Accuracy} \\ \cline{1-7}
Metric     & {\bf Specificity}         & Precision      & F1             & {\bf Sensitivity}         &  Precision      & F1             &                           \\ \hline
COVIDNet-CXR& 83.42$\pm$ 15.39 &	69.73$\pm$ 10.34 &	74.45$\pm$ 8.86 &	61.82$\pm$ 22.44 &	79.50$\pm$ 11.47 &	65.64$\pm$ 15.90 &	72.62$\pm$ 7.6 \\ \hline
COVID-CAPS & 65.09$\pm$ 10.51 &	71.72$\pm$ 5.57 &	67.52$\pm$5.29 &	73.31$\pm$9.74 &	68.40$\pm$5.13 &	70.20$\pm$4.31 &	69.20$\pm$3.61 \\ \hline

\hline\hline
 With seg.      & 80.57$\pm$8.72     &  78.68$\pm$6.57  &  78.97$\pm$3.20   &  76.80$\pm$10.15  &  80.70$\pm$5.56 &  78.01$\pm$4.29  &  78.69$\pm$3.00  \\ \hline
 FuCiTNet & 82.63$\pm$6.61  &	\textbf{79.94$\pm$4.28}  & 81.05$\pm$3.44 &	\textbf{78.91$\pm$5.88} & 82.43$\pm$5.43 & 80.37$\pm$3.16 & 80.77$\pm$3.15 \\ \hline
COVID-SDNet    & {\bf 85.20$\pm$5.38}   & 78.88$\pm$3.89 & {\bf 81.75$\pm$2.74} & 76.80$\pm$6.30 & {\bf  84.23$\pm$4.59} & {\bf  80.07$\pm$0.04} & {\bf 81.00$\pm$2.87} \\ \hline
\end{tabular}}
\caption{Results of the baseline classification model with segmentation, COVID-SDNet, retrained COVIDNet-CXR-A and retrained  COVID-CAPS. Only three levels of severity are considered, Mild, Moderate and Severe.}
\label{tab:SinNormal}
\end{table*}

\subsection{Analysis of the impact of  Normal-PCR+}

To analyze the impact of Normal-PCR+ class on COVID-19 classification, we trained and evaluated the baseline model, FuciTNet,  COVID-SDNet classification stage, COVIDNet-CXR-A and COVID-CAPS,  on  COVIDGR-1.0 by eliminating Normal-PCR+. 
The results are summarized in Table \ref{tab:SinNormal}.

Overall, all the approaches systematically provide better results when eliminating Normal-PCR+ from the training and test processes, including COVIDNet-CXR-A and COVID-CAPS. In particular, COVID-SDNet still represents the best and most stable approach.

\subsection*{Analysis per  severity level}
A further analysis of the accuracy at the level of each severity degree
(see Table \ref{tab:severWOnormal}) demonstrates that eliminating Normal-PCR+ decreases the accuracy in Mild and Moderate severity levels by 15.8\% and 1.52\% respectively. 

 \begin{table}[h]
 \centering
\begin{tabular}{|l|c|}
\hline
S (Severity level)      &  accuracy (S)($\%$)  \\\hline \hline 
  Mild & 46.00 $\pm$ 7.10 \\
  Moderate & 85.38 $\pm$ 1.85 \\
  Severe & 97.22 $\pm$ 1.86 \\\hline
\end{tabular}
\caption{Results of COVID-SDNet by severity level without considering Normal-PCR+.}
\label{tab:severWOnormal}
\end{table}

These results show that although Normal-PCR+ is the hardest level to predict, its presence improves the accuracy of lower severity levels, especially Mild level.

\begin{figure*}[t]
\centering
\subfloat[Original Positive (Mild)]{\includegraphics[ width=0.3\textwidth] {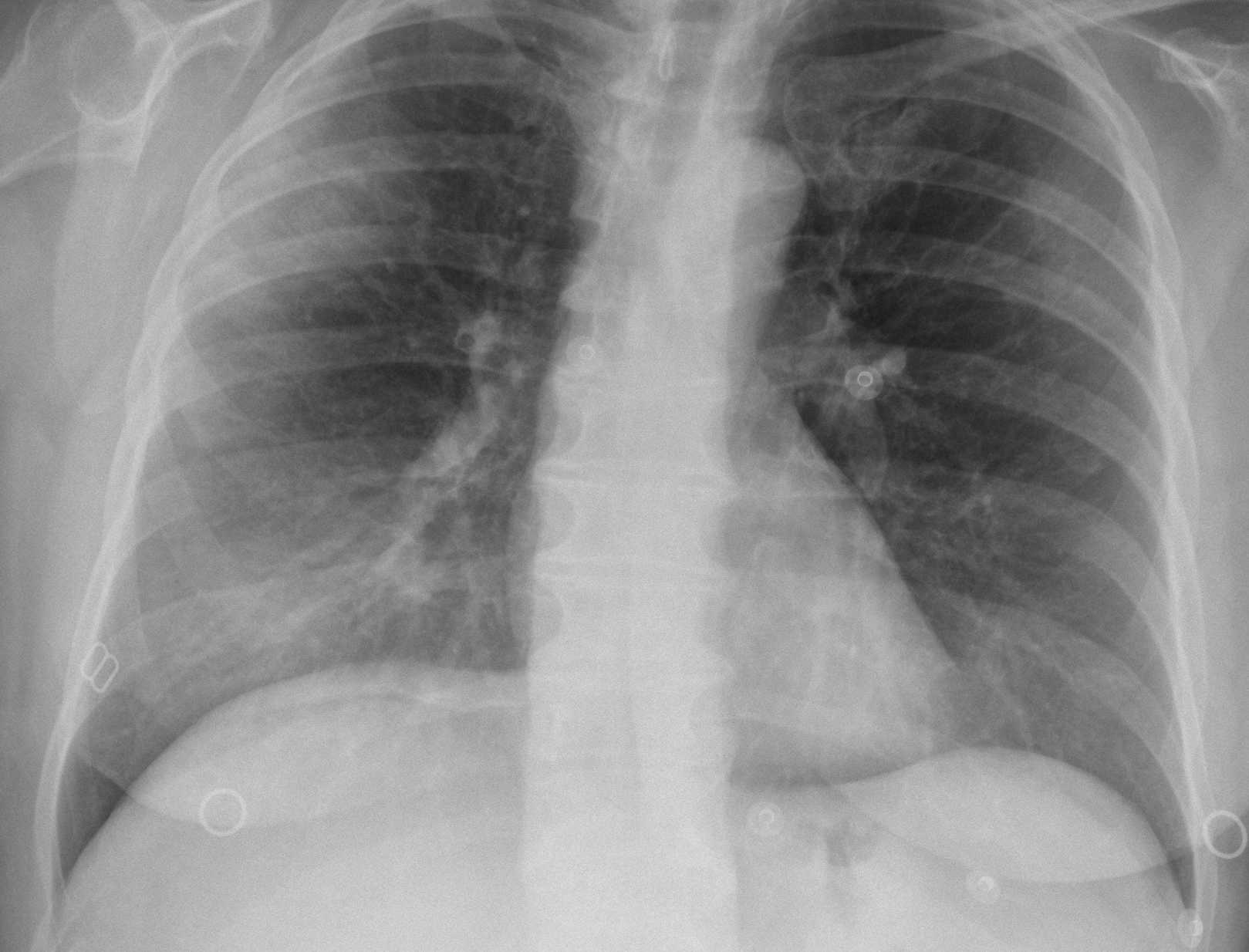}}
\hspace{0.1cm}
\subfloat[why  positive]{\includegraphics[ width=0.3\textwidth] {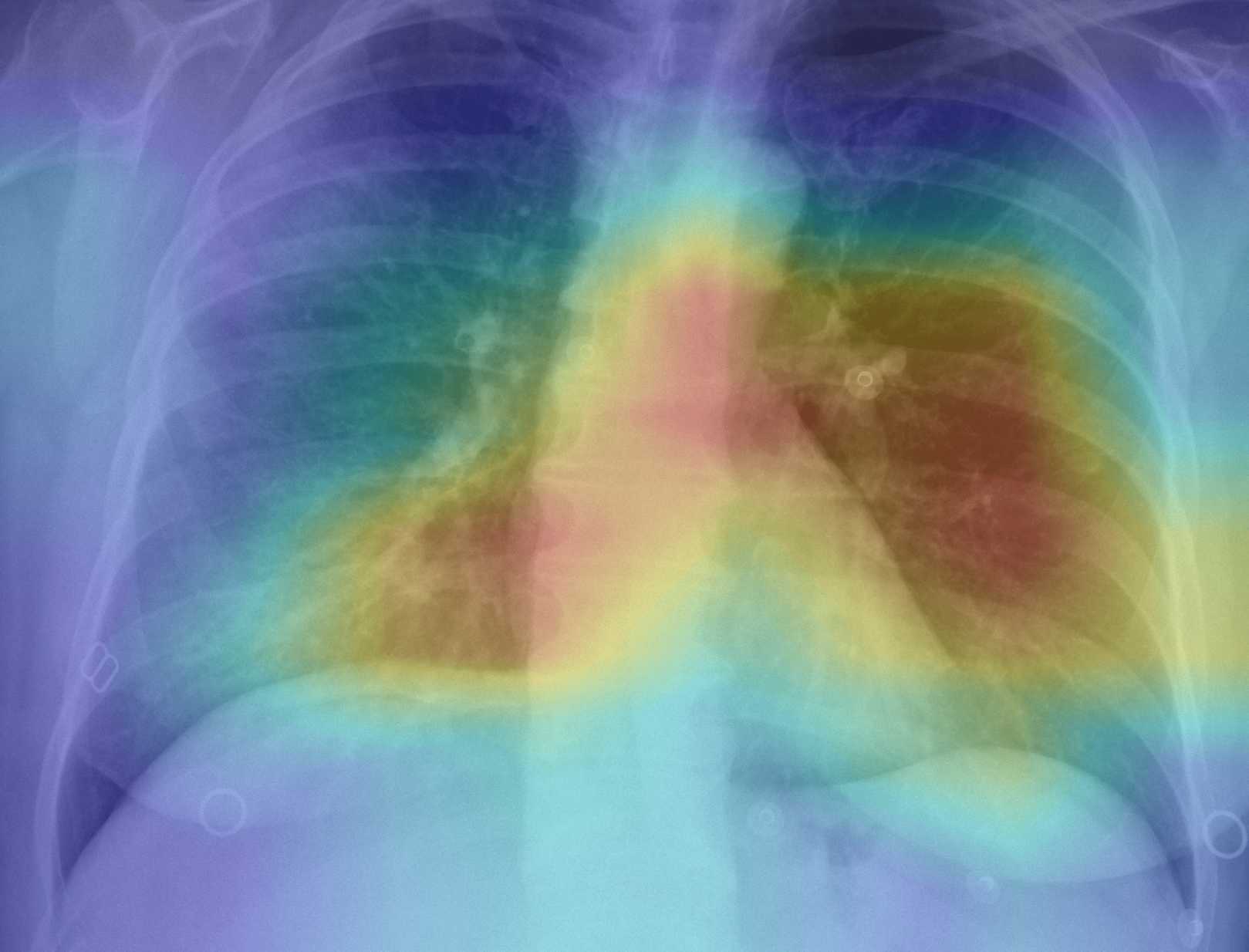}}
\hspace{0.1cm}
\subfloat[why negative]{\includegraphics[ width=0.3\textwidth] {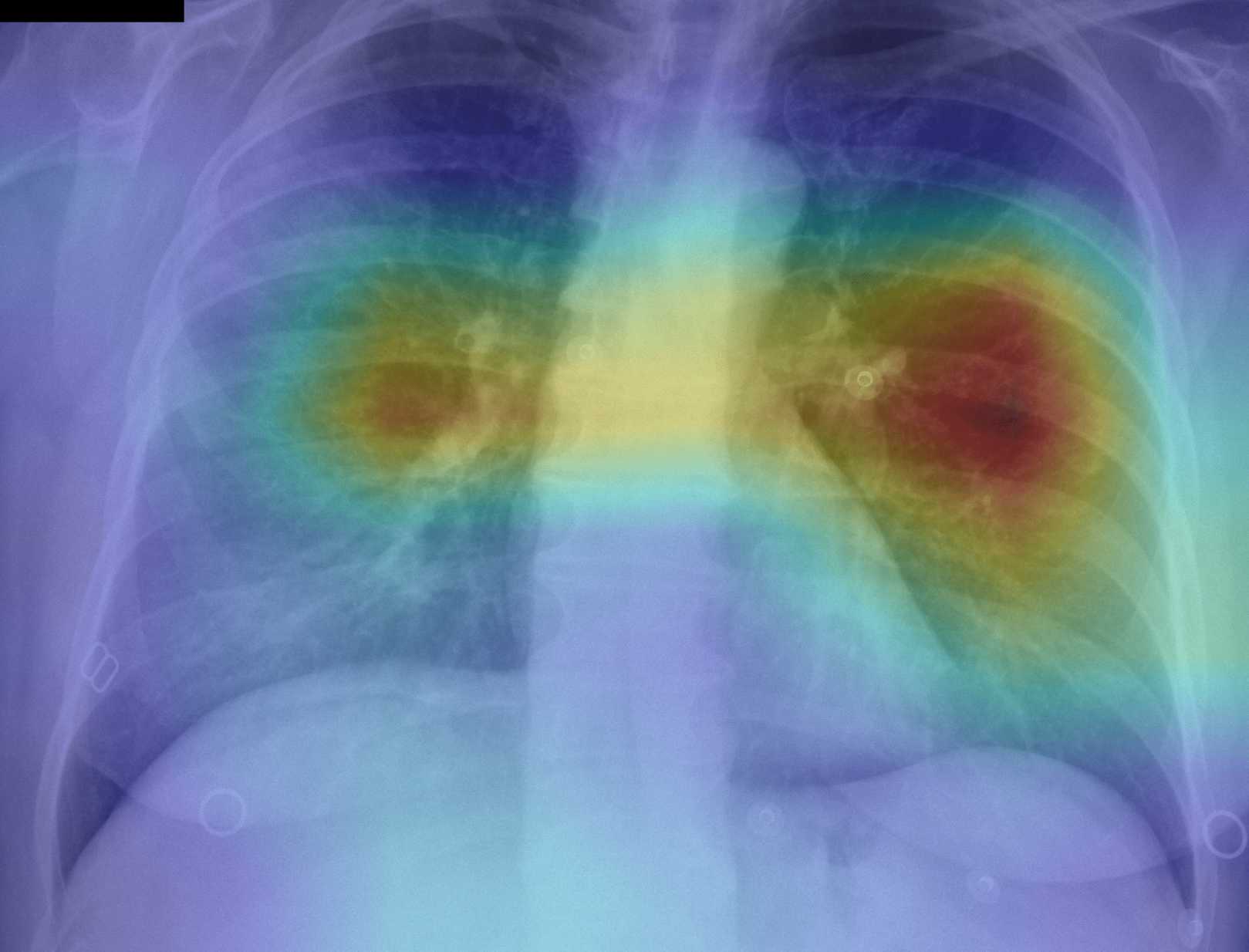}}
\caption{Heatmap showing the parts of the input image that triggered the positive prediction (b) and counterfactual explanation (c)}
\label{fig:6}
\end{figure*}

\begin{figure*}[t]
\centering
\subfloat[Original Positive (Moderate)]{\includegraphics[ width=0.3\textwidth] {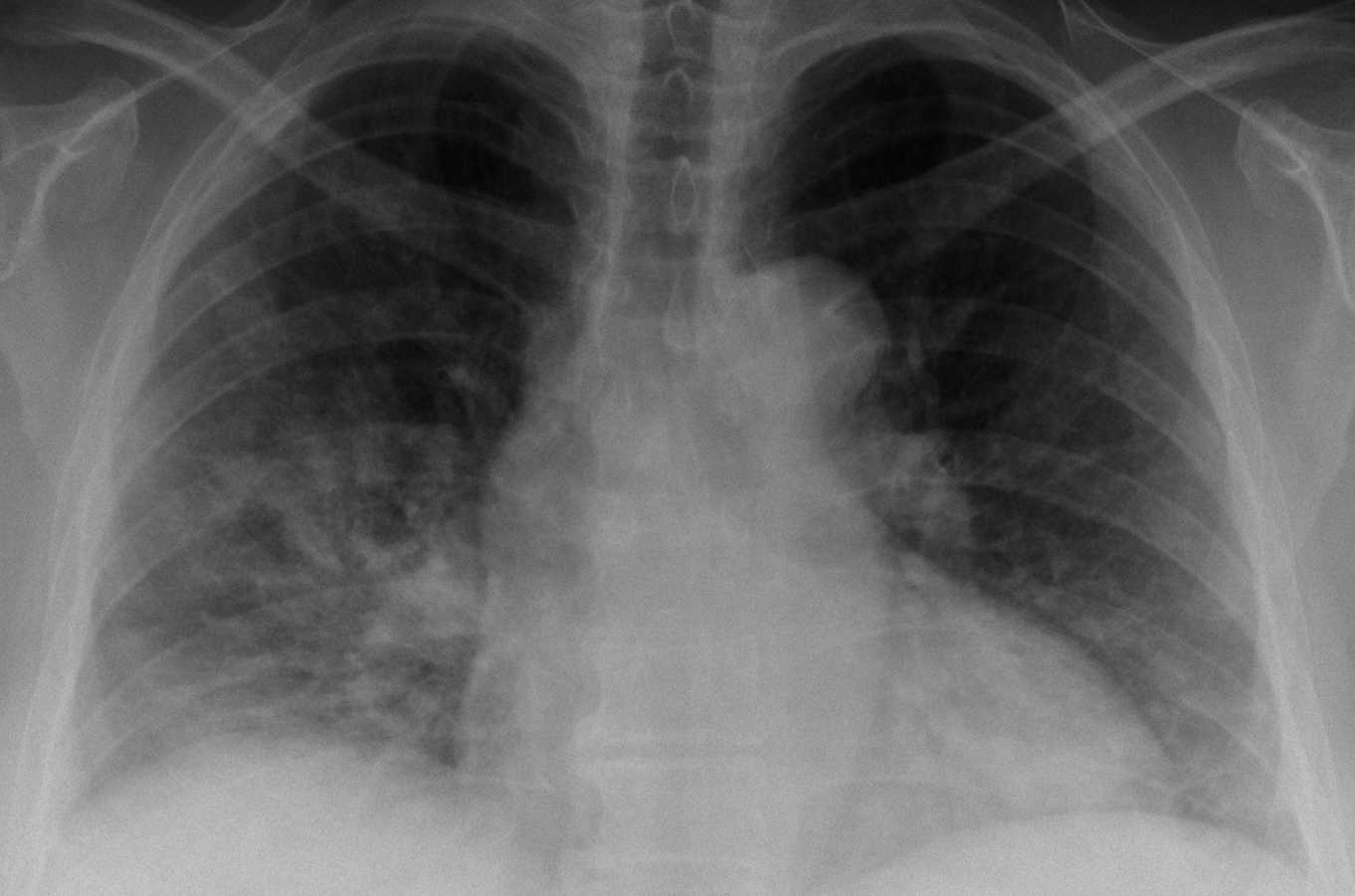}}
\hspace{0.1cm}
\subfloat[why positive]{\includegraphics[width=0.3\textwidth] {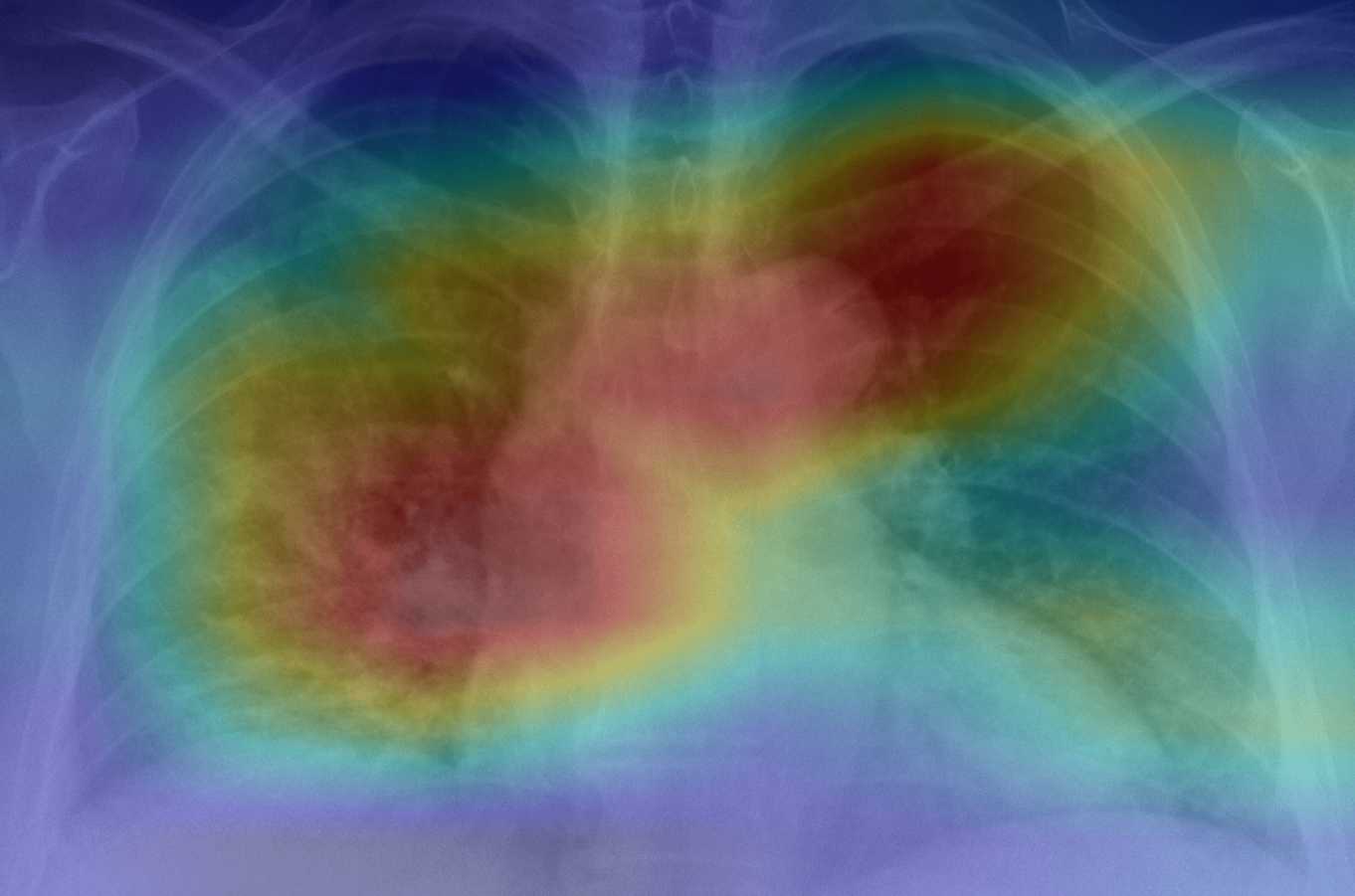}}
\hspace{0.1cm}
\subfloat[why negative]{\includegraphics[ width=0.3\textwidth] {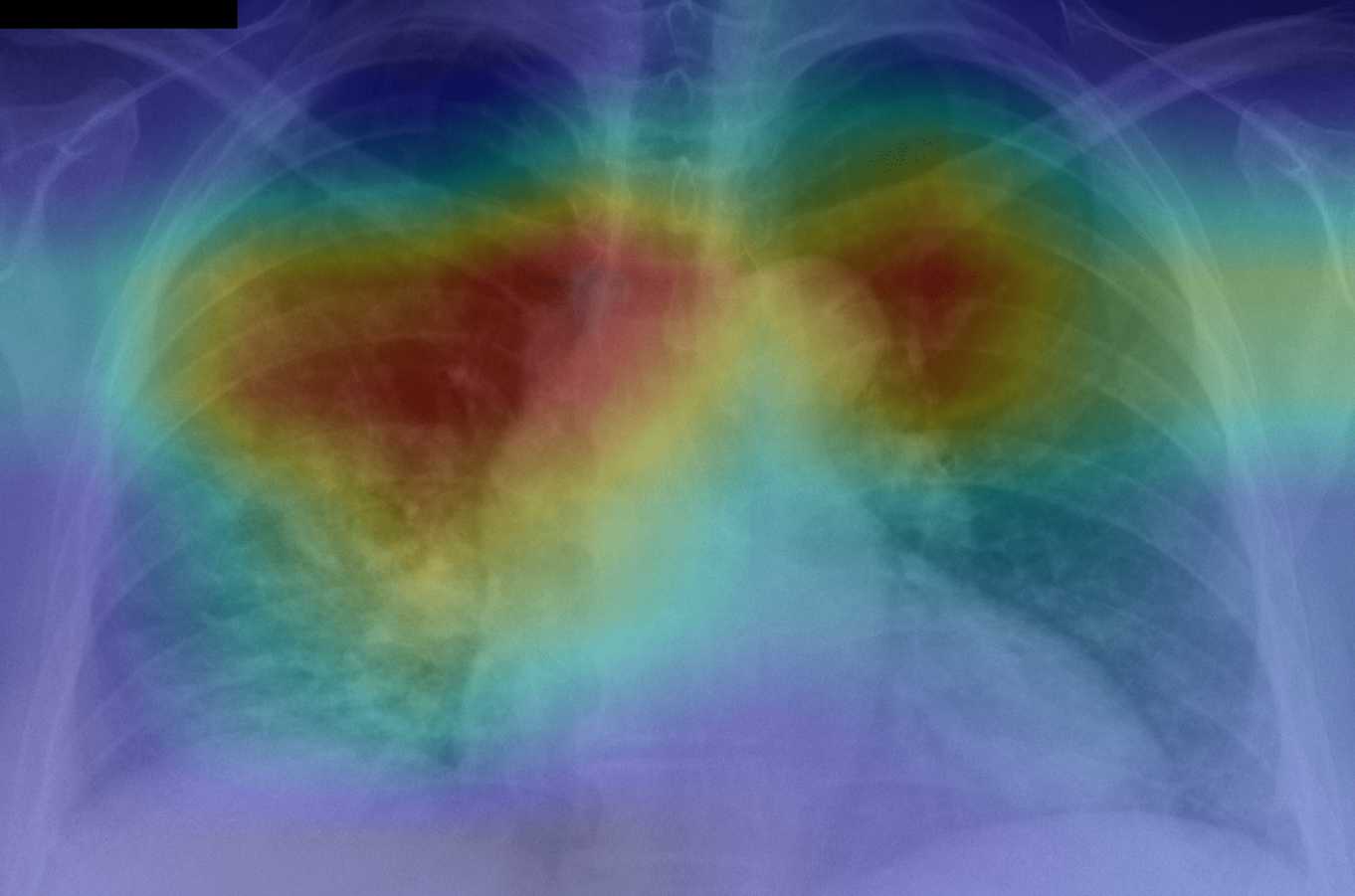}}
\caption{Heatmap showing the parts of the input image that triggered the positive prediction (b) and counterfactual explanation (c)}
\label{fig:7}
\end{figure*}

\begin{figure*}[t]
\centering
\subfloat[Original Positive (Severe)]{\includegraphics[ width=0.3\textwidth] {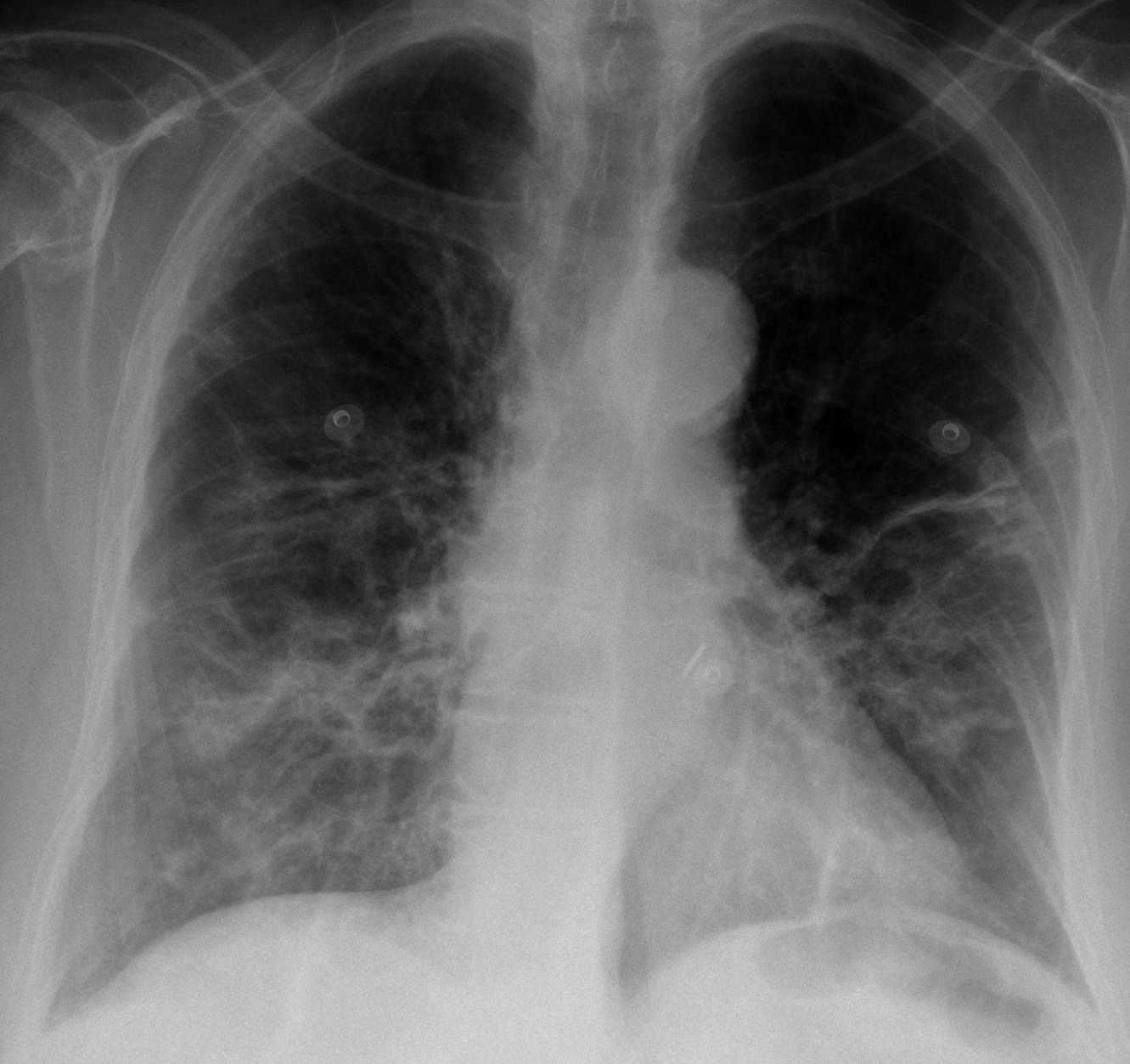}}
\hspace{0.1cm}
\subfloat[why positive]{\includegraphics[ width=0.3\textwidth] {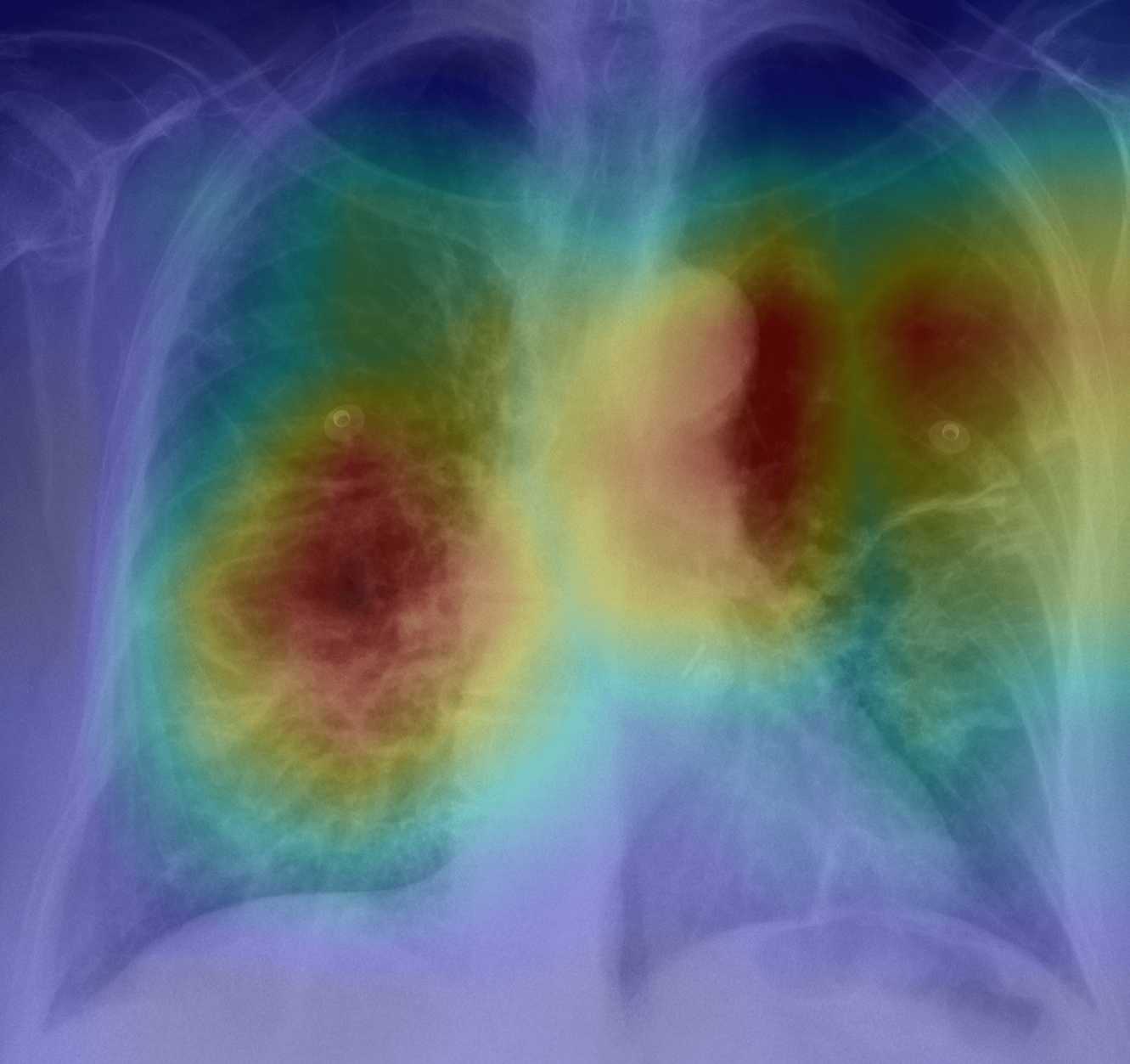}}
\hspace{0.1cm}
\subfloat[why negative]{\includegraphics[ width=0.3\textwidth] {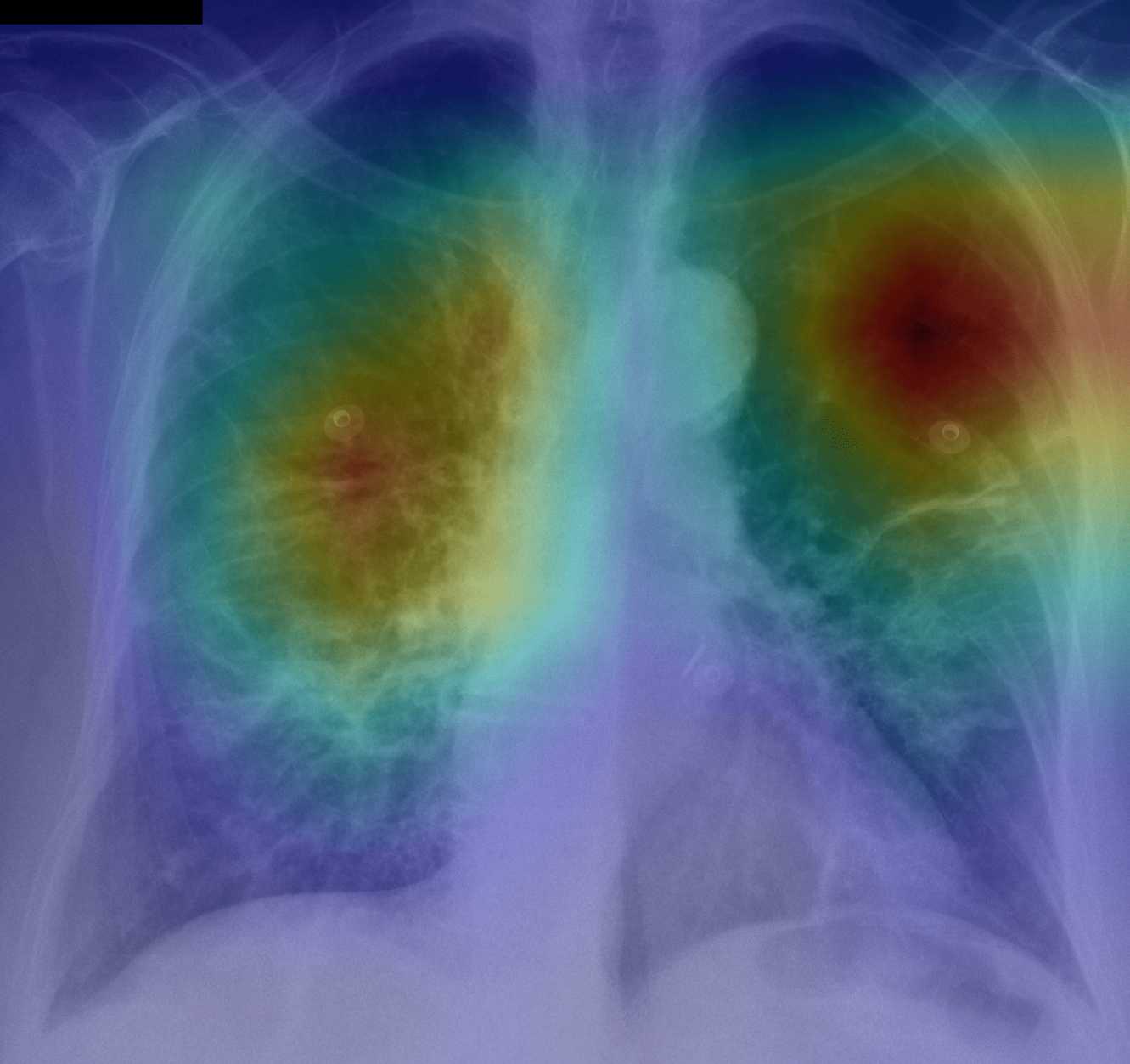}}
\caption{Heatmap showing the parts of the input image that triggered the positive prediction (b) and counterfactual explanation (c)}
\label{fig:8}
\end{figure*}

\begin{figure*}[t]
\centering
\subfloat[Original Negative]{\includegraphics[ width=0.32\textwidth] {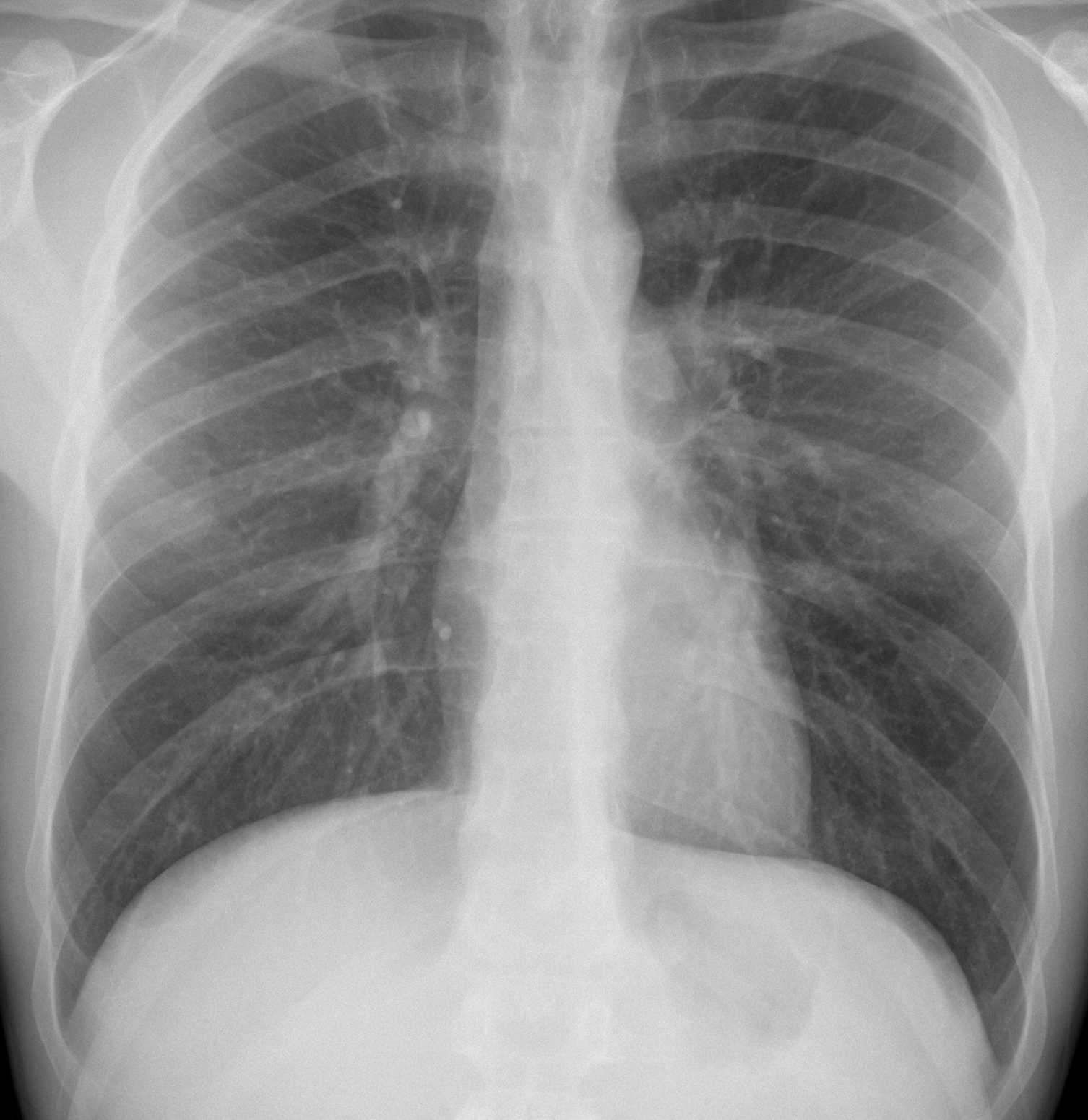}}
\hspace{0.1cm}
\subfloat[why positive]{\includegraphics[ width=0.32\textwidth] {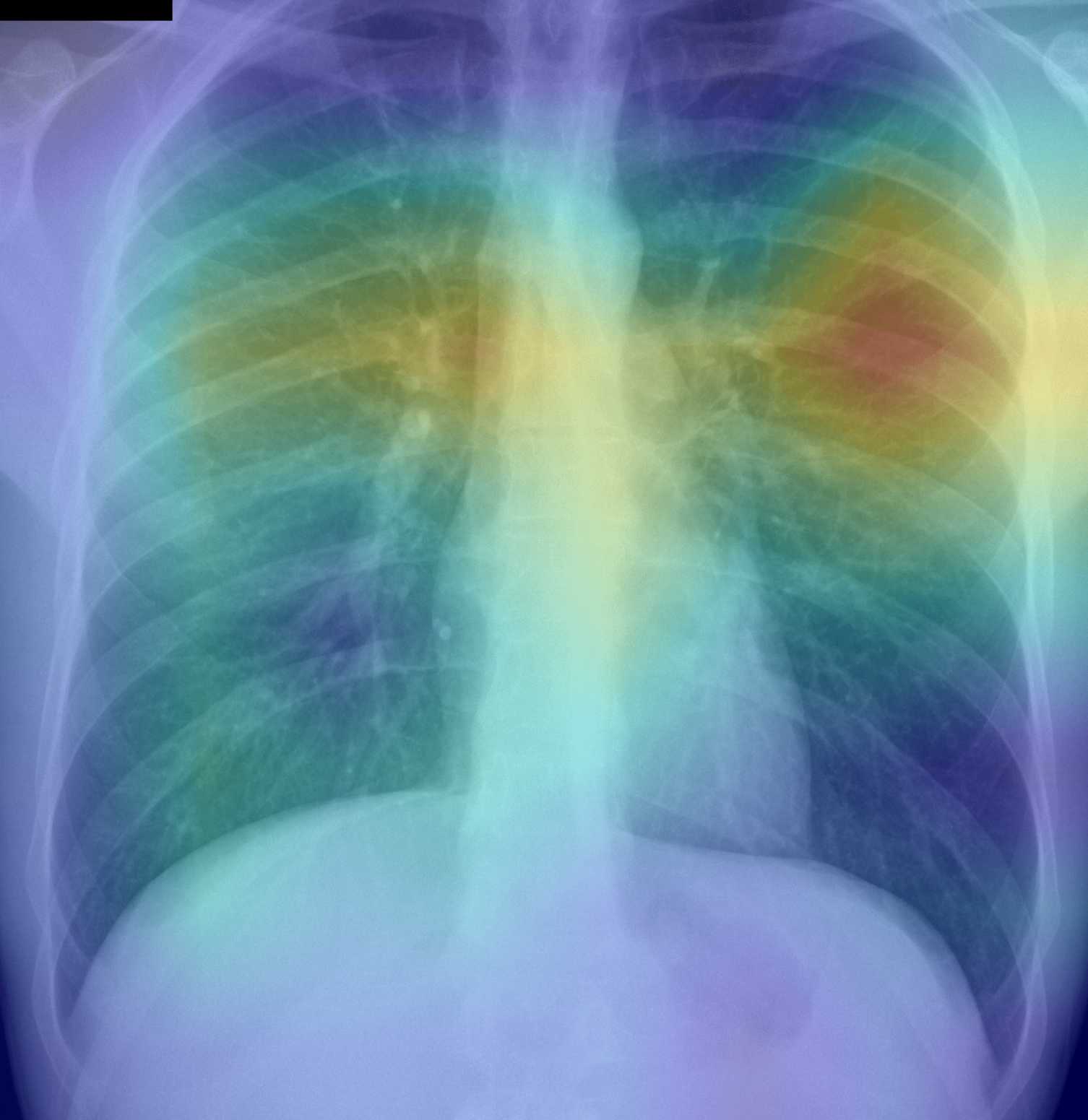}}
\hspace{0.1cm}
\subfloat[why negative]{\includegraphics[ width=0.32\textwidth] {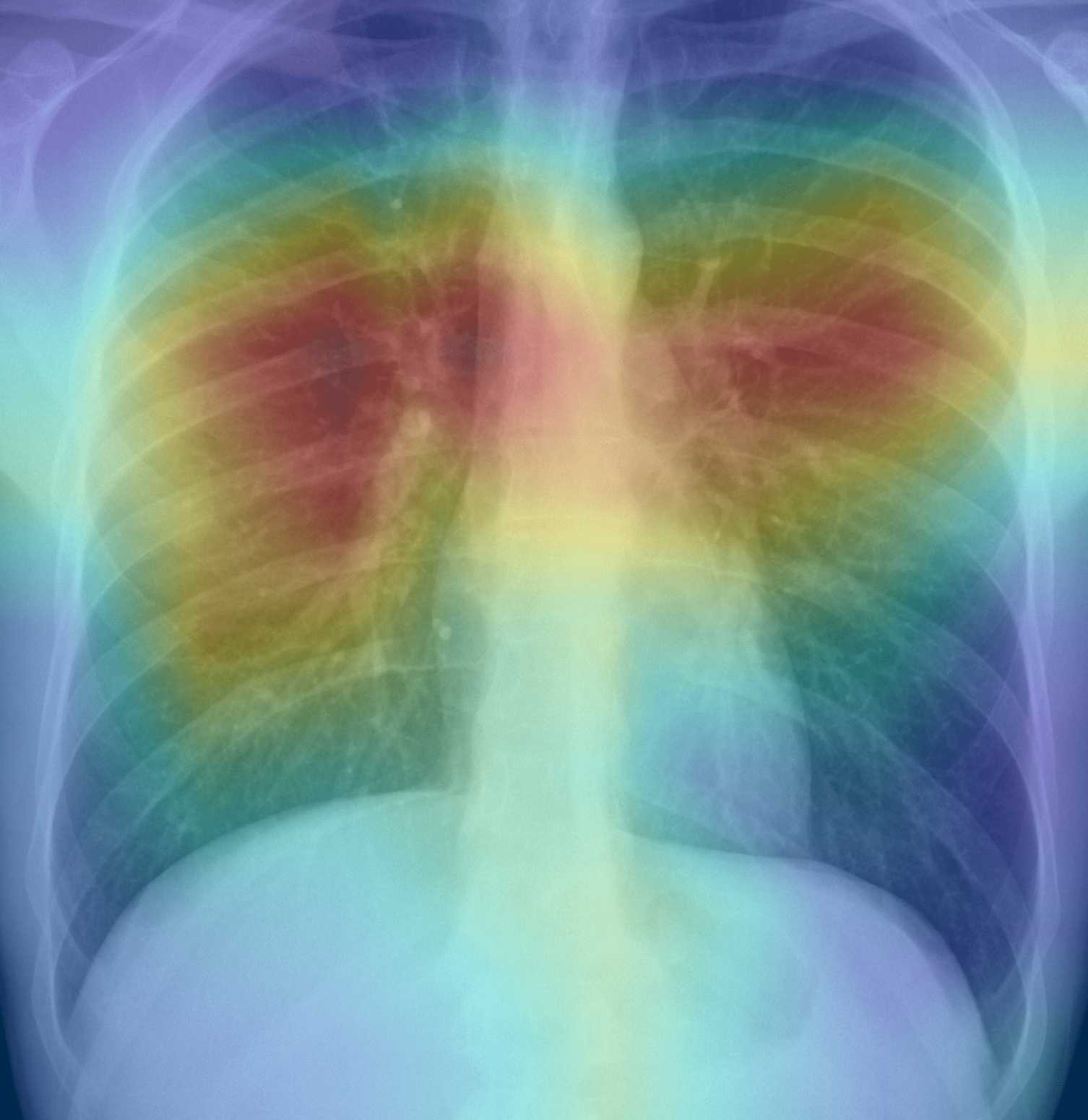}}
\caption{Heatmap that explains the parts of the input image that triggered the counterfactual explanation (b) and the negative actual prediction (c).}
\label{fig:5}
\end{figure*}

\section{Inspection of model's decision}
\label{sec:inspection}
Automatic DL diagnosis systems alone are not mature yet to replace expert radiologists. To help clinician  making decisions, these tools must be interpretable so that  clinicians can decide whether to trust the model or not (\cite{arrieta2020explainable}). We inspect what led our model make a decision by showing the regions of the input image that triggered that decision along with its counterfactual explanation by showing the parts that explain the opposite class. We adapted Grad-CAM method (\cite{selvaraju2017grad}) to explain the decision of the negative and positive class.

 Fig.  \ref{fig:6}, \ref{fig:7} and \ref{fig:8} show (a) the  original CXR image, (b) visual explanation by means of a heat-map that highlights the regions/pixels which led the model to output the actual prediction and (c) its counterfactual explanation using a heat-map that highlights the regions/pixels which had the highest impact on predicting the opposite class.  Higher intensity in the heat-map indicates higher importance of the corresponding pixel in the decision. The larger higher intensity areas in the heat-map determine the final class. However, Fig.  \ref{fig:5}(b) represents first the counterfactual explanation and Fig.  \ref{fig:5}(c) represents the  explanation of the  actual decision.

 As expected, negative and positive interpretations are complementary, i.e, areas which triggered the correct decision  are opposite, in most cases, to the areas that triggered the decision towards negative. In  CXR images with different severity levels, the heat-maps correctly point out opaque regions due to different levels of infiltrates, consolidations and  also to osteoarthritis.
 
 In particular, in Fig.  \ref{fig:6}(b), the  red areas in the right lung points out a region with infiltrates and also   osteoarthritis in the spine region. Fig.  \ref{fig:7}  (b) correctly shows moderate infiltrates in the right lower and lower-middle lung fields in addition to a dilation of ascending aorta and aortic arch (red color in the center). Fig.  \ref{fig:6}(c) shows  normal upper-middle fields of both lungs (less important on the left due to aortic dilation). Fig.  \ref{fig:8}(b) indicates an important bilateral pulmonary involvement with consolidations.

As it can be observed in Fig.  \ref{fig:5}(c), the  explanation of the negative class correctly highlights a symmetric bilateral pattern that occupies a larger  lung volume especially in regions with high density. In fact, a very similar pattern is shown in the counterfactual explanation of the positive class in  Fig.  \ref{fig:6}(c), \ref{fig:7}(c) and \ref{fig:8}(c).

\section{Conclusions}
\label{sec:conclusions}
This paper introduced a  dataset, named COVIDGR-1.0, with high clinical value. COVIDGR-1.0 includes the  four main COVID severity levels identified by a recent radiological study  (\cite{wong2020frequency}). We  proposed a methodology, called COVID-SDNet, that combines segmentation, data-augmentation and data transformation. The obtained results show the high generalization capacity of COVID-SDNet, specially on severe and moderate levels as they include important visual features. The existence of  few or none visual features in Mild and Normal-PCR+ reduces the opportunities for improvement. 

As main conclusions, we must highlight that COVID-SDNet can be used in a triage system to detect especially moderate and severe patients. Finally, we must also mention that  more robust and accurate triage system can be built by fusing our approach  with other approaches such as the one proposed in (\cite{CohenAnnot}).

As future work, we are working on enriching COVIDGR-1.0 with more CXR images coming from different hospitals. We are  planning to explore the use of additional clinical information along with  CXR images to improve the prediction performance.

\section*{Acknowlegments}
This work was supported by the project DeepSCOP-Ayudas Fundación BBVA a Equipos de Investigación Científica en Big Data 2018, COVID19\_RX-Ayudas Fundación BBVA a Equipos de Investigación Científica SARS-CoV-2 y COVID-19 2020, and
the Spanish Ministry of Science and Technology under the project TIN2017-89517-P. S.  Tabik was supported by the Ramon y Cajal Programme (RYC-2015-18136). A. Gómez-Ríos was supported by the FPU Programme FPU16/04765. D. Charte was supported by the FPU Programme FPU17/04069. J. Su{\'a}rez was supported by the FPU Programme FPU18/05989. E.G was supported by the European Research Council (ERC Grant agreement 647038 [BIODESERT])

\section*{Ethics}
This project is approved by the Provincial Research Ethics Committee of Granada.




%

\bibliographystyle{plain}



\end{document}